\input harvmac 
\input epsf.tex

\overfullrule=0mm

\newcount\figno
\figno=0
\def\fig#1#2#3{
\par\begingroup\parindent=0pt\leftskip=1cm\rightskip=1cm\parindent=0pt
\baselineskip=11pt
\global\advance\figno by 1
\midinsert
\epsfxsize=#3
\centerline{\epsfbox{#2}}
\vskip 12pt
{\bf Fig.\the\figno:} #1\par
\endinsert\endgroup\par
}
\def\figlabel#1{\xdef#1{\the\figno}}
\def\encadremath#1{\vbox{\hrule\hbox{\vrule\kern8pt\vbox{\kern8pt
\hbox{$\displaystyle #1$}\kern8pt}
\kern8pt\vrule}\hrule}}

\def\wrt{with respect to\ }

\def\IR{\relax{\rm I\kern-.18em R}}
\font\cmss=cmss10 \font\cmsss=cmss10 at 7pt

\font\cmss=cmss10 \font\cmsss=cmss10 at 7pt
\def\IZ{\relax\ifmmode\mathchoice
{\hbox{\cmss Z\kern-.4em Z}}{\hbox{\cmss Z\kern-.4em Z}}
{\lower.9pt\hbox{\cmsss Z\kern-.4em Z}}
{\lower1.2pt\hbox{\cmsss Z\kern-.4em Z}}\else{\cmss Z\kern-.4em Z}\fi}
\def\IN{\relax{\rm I\kern-.18em N}}
\def\b{\circ}
\def\n{\bullet}

\def\gbbbb{\Gamma_4^{\hbox{$\scriptstyle \b \b$}\kern -8.2pt
\raise -4pt \hbox{$\scriptstyle \b \b$}}}
\def\gnnnn{\Gamma_4^{\hbox{$\scriptstyle \n \n$}\kern -8.2pt  
\raise -4pt \hbox{$\scriptstyle \n \n$}}}
\def\gnnnnnn{\Gamma_6^{\hbox{$\scriptstyle \n \n \n$}\kern -12.3pt
\raise -4pt \hbox{$\scriptstyle \n \n \n$}}}
\def\gbbbbbb{\Gamma_6^{\hbox{$\scriptstyle \b \b \b$}\kern -12.3pt
\raise -4pt \hbox{$\scriptstyle \b \b \b$}}}
\def\gbbbbc{\Gamma_{4\, c}^{\hbox{$\scriptstyle \b \b$}\kern -8.2pt
\raise -4pt \hbox{$\scriptstyle \b \b$}}}
\def\gnnnnc{\Gamma_{4\, c}^{\hbox{$\scriptstyle \n \n$}\kern -8.2pt
\raise -4pt \hbox{$\scriptstyle \n \n$}}}
\def\Rbud#1{{\cal R}_{#1}^{-\kern-1.5pt\blacktriangleright}}
\def\Rleaf#1{{\cal R}_{#1}^{-\kern-1.5pt\vartriangleright}}
\def\Rbudb#1{{\cal R}_{#1}^{\circ\kern-1.5pt-\kern-1.5pt\blacktriangleright}}
\def\Rleafb#1{{\cal R}_{#1}^{\circ\kern-1.5pt-\kern-1.5pt\vartriangleright}}
\def\Rbudn#1{{\cal R}_{#1}^{\bullet\kern-1.5pt-\kern-1.5pt\blacktriangleright}}
\def\Rleafn#1{{\cal R}_{#1}^{\bullet\kern-1.5pt-\kern-1.5pt\vartriangleright}}
\def\Wleaf#1{{\cal W}_{#1}^{-\kern-1.5pt\vartriangleright}}
\def\Cleaf{{\cal C}^{-\kern-1.5pt\vartriangleright}}
\def\Cbud{{\cal C}^{-\kern-1.5pt\blacktriangleright}}
\def\Crleaf{{\cal C}^{-\kern-1.5pt\circledR}}


\Title{\vbox{\hsize=3.truecm \hbox{SPhT/03-029}}}
{{\vbox {
\bigskip
\centerline{Geodesic Distance in Planar Graphs}
}}}
\bigskip
\centerline{J. Bouttier\foot{bouttier@spht.saclay.cea.fr}, 
P. Di Francesco\foot{philippe@spht.saclay.cea.fr} and
E. Guitter\foot{guitter@spht.saclay.cea.fr}}
\medskip
\centerline{ \it Service de Physique Th\'eorique, CEA/DSM/SPhT}
\centerline{ \it Unit\'e de recherche associ\'ee au CNRS}
\centerline{ \it CEA/Saclay}
\centerline{ \it 91191 Gif sur Yvette Cedex, France}
\bigskip
\noindent We derive the exact generating function for planar maps 
(genus zero fatgraphs) with vertices of arbitrary even valence and 
with two marked points at a fixed geodesic distance. This is done 
in a purely combinatorial way based on a bijection with decorated 
trees, leading to a recursion relation on the geodesic distance. 
The latter is solved exactly in terms of discrete soliton-like 
expressions, suggesting an underlying integrable structure. We 
extract from this solution the fractal dimensions at the various 
(multi)-critical points, as well as the precise scaling forms of 
the continuum two-point functions and the probability distributions 
for the geodesic distance in (multi)-critical random surfaces. 
The two-point functions are shown to obey differential equations 
involving the residues of the KdV hierarchy.
\Date{03/03}

\nref\TUT{W. Tutte, 
{\it A Census of planar triangulations}
Canad. Jour. of Math. {\bf 14} (1962) 21-38;
{\it A Census of Hamiltonian polygons}
Canad. Jour. of Math. {\bf 14} (1962) 402-417;
{\it A Census of slicings}
Canad. Jour. of Math. {\bf 14} (1962) 708-722;
{\it A Census of Planar Maps}, Canad. Jour. of Math. 
{\bf 15} (1963) 249-271.}
\nref\BIPZ{E. Br\'ezin, C. Itzykson, G. Parisi and J.-B. Zuber, {\it Planar
Diagrams}, Comm. Math. Phys. {\bf 59} (1978) 35-51.}
\nref\DGZ{P. Di Francesco, P. Ginsparg
and J. Zinn--Justin, {\it 2D Gravity and Random Matrices},
Physics Reports {\bf 254} (1995) 1-131.}
\nref\EY{B. Eynard, {\it Random Matrices}, Saclay Lecture Notes (2000),
available at {\sl http://www-spht.cea.fr/lectures\_notes.shtml} }
\nref\KPZ{V.G. Knizhnik, A.M. Polyakov and A.B. Zamolodchikov, {\it Fractal Structure of
2D Quantum Gravity}, Mod. Phys. Lett.
{\bf A3} (1988) 819-826; F. David, {\it Conformal Field Theories Coupled to 2D Gravity in the
Conformal Gauge}, Mod. Phys. Lett. {\bf A3} (1988) 1651-1656; J.
Distler and H. Kawai, {\it Conformal Field Theory and 2D Quantum Gravity},
Nucl. Phys. {\bf B321} (1989) 509-527.}
\nref\KKMW{H. Kawai, N. Kawamoto, T. Mogami and Y. Watabiki, {\it Transfer Matrix 
Formalism for Two-Dimensional Quantum Gravity and Fractal Structures of Space-time}, 
Phys. Lett. B {\bf 306} (1993) 19-26.}
\nref\AW{J. Ambj\o rn and Y. Watabiki, {\it Scaling in quantum gravity},
Nucl.Phys. {\bf B445} (1995) 129-144.}
\nref\AJW{J. Ambj\o rn, J. Jurkiewicz and Y. Watabiki, 
{\it On the fractal structure of two-dimensional quantum gravity},
Nucl.Phys. {\bf B454} (1995) 313-342.}
\nref\CS{P. Chassaing and G. Schaeffer, {\it Random Planar Lattices and 
Integrated SuperBrownian Excursion}, preprint (2002), to appear in 
Probability Theory and Related Fields, math.CO/0205226.}
\nref\SCH{G. Schaeffer, {\it Bijective census and random 
generation of Eulerian planar maps}, Electronic
Journal of Combinatorics, vol. {\bf 4} (1997) R20; see also
G. Schaeffer, {\it Conjugaison d'arbres
et cartes combinatoires al\'eatoires} PhD Thesis, Universit\'e 
Bordeaux I (1998).}
\nref\CENSUS{J. Bouttier, P. Di Francesco and E. Guitter, {\it Census of planar
maps: from the one-matrix model solution to a combinatorial proof},
Nucl. Phys. {\bf B645}[PM] (2002) 477-499.}
\nref\BMS{M. Bousquet-M\'elou and G. Schaeffer,
{\it Enumeration of planar constellations}, Adv. in Applied Math.,
{\bf 24} (2000) 337-368.}
\nref\JM{M. Jimbo and T. Miwa, {\it Solitons and infinite dimensional Lie
algebras}, Publ. RIMS, Kyoto Univ. {\bf 19} No. 3 (1983) 943-1001, 
eq.(2.12).}
\nref\STAU{M. Staudacher, {\it The Yang-Lee Edge Singularity on a
Dynamical Planar Random Surface}, Nucl. Phys. {\bf B336} (1990) 349-362.}
\nref\GD{I. Gelfand and L. Dikii, {\it Fractional powers of operators and
Hamiltonian systems}, Funct. Anal. Appl. {\bf 10:4} (1976) 13.}
\nref\KY{N. Kawamoto and K. Yotsuji, {\it Numerical study for the c-dependence 
of fractal dimension in two-dimensional quantum gravity}, 
Nucl.Phys. {\bf B644} (2002) 533-567.}
\nref\BFG{J. Bouttier, P. Di Francesco and E. Guitter,
{\it Counting colored Random Triangulations}, 
Nucl.Phys. {\bf B641} (2002) 519-532.}
\nref\CONC{M. Bousquet-M\'elou and G. Schaeffer, {\it The degree distribution
in bipartite planar maps: application to the Ising model}, preprint
math.CO/0211070.}
\nref\NOUSHARD{J. Bouttier, P. Di Francesco and E. Guitter, {\it
Combinatorics of hard particles on planar maps} preprint (2002)
to appear in Nucl. Phys. {\bf B}, cond-mat/0211168.}
\nref\NOUSISING{
J. Bouttier, P. Di Francesco and E. Guitter, work in progress.}
\nref\DEL{J.-F. Delmas {\it Computation of moments for the length of the 
one dimensional ISE support}, preprint (2002), available at
http://cermics.enpc.fr/$\sim$delmas/.}
\nref\DGL{P. Di Francesco and E. Guitter, {\it Critical and multicritical
semi-random $(1+d)$-dimensional lattices and hard objects in $d$
dimensions}, J. Phys. A Math. Gen, {\bf 35} (2002) 897-927.}

\newsec{Introduction}

The study of the statistical properties of random graphs 
is relevant for many problems in physics, such as two-dimensional quantum gravity or
fluid membrane statistics. By random graphs, we here mean graphs {\it embedded} in
a surface of given genus, also known as {\it fatgraphs} or {\it maps}. Indeed, these
are the natural discretization of fluctuating surfaces, on which matter systems may 
be defined. Much is known to this day on the enumeration of maps of fixed genus either by 
combinatorial techniques \TUT\ or by the use of matrix integrals [\xref\BIPZ-\xref\EY]. 
When applied to graphs of large size, these results give rise to various
scaling behaviors depending on the critical universality class at hand, described
in the continuum by the coupling to 2D quantum gravity of some 2D Conformal Field 
Theories (CFT) with central charges $c<1$. In particular, the critical behaviors  
are characterized by the famous KPZ scaling relations \KPZ. As an example, the so-called 
one-matrix model, which enumerates fatgraphs with vertices of arbitrary valence
and of any fixed topology, displays a set of  multicritical points corresponding to the
CFT's with central charges $c(2,2m+1)$, $m=1,2,3,...$. The first value $m=1$ corresponds
to $c=0$, i.e. a model of pure 2D quantum gravity without matter, representing the 
universality class of generic random maps of large size. The higher order multicritical 
points may be reached by weighting vertices according to their valence, and by 
suitably fine-tuning the vertex weights. 

Most results obtained so far concern {\it global} properties of the graphs
which, in the continuum, can be translated into correlation functions integrated 
over the positions of their insertion points on the surfaces. 
Little is known however on more refined properties of random graphs such as 
the dependence of correlators on the distances between their insertion points.
By distance we mean here {\it geodesic distance} along the graph, namely the length
of any shortest path between say two given faces.
These refined properties seem to remain beyond the reach of the standard matrix model 
treatment.  
Still, some results were obtained in Refs.[\xref\KKMW-\xref\AJW], where in particular
combinatorial arguments were used to derive the universal scaling two-point function for
critical genus zero surfaces with two marked points at a fixed geodesic distance. 
However, an explicit form for the two-point function was obtained only in the pure 
gravity case and close to the critical point. 

More recently, using a completely different combinatorial approach, the same result was recovered
in Ref.\CS\ in the case of random tetravalent planar maps. This work relies
on the existence of a bijection between planar tetravalent maps and 
rooted trees with vertices labelled by the geodesic distance to the root. 
The construction however is specific to the tetravalent case, and does not provide
a closed form for the discrete solution either. 

In this note, we address the general question of the enumeration of planar maps
with two marked faces at a fixed geodesic distance. In the case of maps with inner 
vertices of arbitrary even valence, we derive explicit expressions for the generating
function $G_n$ of ``two-leg diagrams", namely planar maps with two distinguished univalent
vertices (legs) whose adjacent faces lie at a geodesic distance $n$ of one-another, and
with weights $g_i$ per inner $2i$-valent vertex, $i=1,2,3,...$. 
These results are obtained via the use of a more general bijection between two-leg
diagrams and decorated trees first found in Ref.\SCH\ and extended
in Ref.\CENSUS, which we adapt so as to keep
track of the geodesic distance between the two legs. This leads to a simple algebraic recursion
relation on $n$, which allows to derive a closed expression  for the solution $G_n$.
We give several explicit expressions depending on the maximum valence
of the vertices of the graph. With these solutions in hand, we easily derive
the continuum scaling two-point functions corresponding to the various
multicritical points. This allows to recover the results of Refs.\AW\ and \CS\ in the
generic critical case, and also provides the generalization to all higher
order multicritical points. We also extract critical exponents such as the fractal dimension
in various approaches to the multicritical points, as well as probability distributions 
for the (rescaled) geodesic distance in multicritical planar maps of fixed but large size.  

The paper is organized as follows. 
In Sect.2 we detail the general correspondence between two-leg diagrams
and decorated trees. Sect.3 is devoted to the incorporation of the geodesic distance
$n$ between the two legs in this setting and leads to a general recursion relation on
$n$. This recursion relation is solved first in the tetravalent case (Sect.4.1),
and exploited to derive a critical fractal dimension $d_F=4$ (Sect.4.2) 
and to recover the known continuum two-point function for critical surfaces (Sect.4.3).
We also use our explicit solution to derive in Sect.4.4 the probability distribution for
rescaled geodesic distances in two-leg diagrams of fixed but large size. 
The general case of maps with valences up to $2(m+1)$ is solved in Sect.5, where we
discover a remarkable connection with $m$-soliton solutions of the KP hierarchy.
This general solution is applied in Sect.6 to the so-called ``hard dimer" model, which 
displays the first higher order critical point (tricritical point). In Sect.6.1 we detail 
the solution and display in particular the first few $G_n$'s. This leads to the computation
in Sect.6.2  of the fractal dimension $d_F=6$ and the continuum two-point function 
for tricritical surfaces, corresponding to a generic approach to the tricritical point.
We also discuss the case of a non-generic approach along the critical line, which
displays another fractal dimension $d_F=4$ and a different two-point scaling function.
We finally obtain in Sect.6.3 the probability distribution for rescaled 
geodesic distances in tricritical surfaces of fixed but large size.
All these results are extended to the case of the $m$-th order multicritical point
in Sect.7, where we obtain the fractal dimension $d_F=2(m+1)$ (Sect.7.1), the generic
continuum multicritical two-point function (Sect.7.2) and the corresponding
probability distribution (Sect.7.3).
We gather a few concluding remarks in Sect.8, namely the relation to matrix models (Sect.8.1),
the extension to two-point functions of other (less relevant) operators (Sect.8.2), some
generalizations to other classes of planar graphs such as constellations (Sect.8.3),
as well as a connection to the so-called Integrated SuperBrownian Excursion (ISE)
(Sect.8.4).
We also gather a few technical details in appendices A and B.

\newsec{Planar maps and decorated trees}

There is a general correspondence between planar maps  
and special classes of decorated trees [\xref\SCH,\xref\CENSUS].  
In this paper, we concentrate on the simpler case of
maps with only vertices of {\it even} valence.
In this case, there is a bijection between 
so-called ``two-leg diagrams" and so-called rooted
``blossom" trees as defined below. 
By two-leg diagram, we mean a planar map with two
external legs (i.e. two extra univalent vertices) 
distinguished as in- and out-coming, whereas all inner 
vertices have even valences. As opposed to the definition used in
Ref.\CENSUS, we {\it do not} require that the two external legs lie
in the {\it same} face. By convention, we decide
to always represent the diagram in such a way that the in-coming leg
is adjacent to the external face in the plane.  
On the other hand, by rooted blossom trees we mean
rooted planar trees with two kinds of endpoints called buds and leaves,  
with only inner vertices of even valences and such that  
each inner $2k$-valent vertex is connected to exactly $k-1$ buds.
It is easily seen that any such tree has exactly {\it one} more leaf than bud.

\fig{Illustration of the one-to-one correspondence between two-leg diagrams
and blossom trees. Starting from a two-leg diagram (a), we apply the iterative 
cutting procedure explained in the text, which here requires two turns around
the graph. The edges cut during the $1$st and $2$nd turn are displayed in (b) 
with the corresponding index $1,2$. Each cut edge is replaced by a bud-leaf pair
(c), while the in-coming leg is replaced by a leaf and the out-coming one by a root,
finally leading to a blossom tree (d). Conversely, the matching of buds and leaves
of the blossom tree (d) rebuilds the edges of (a).}{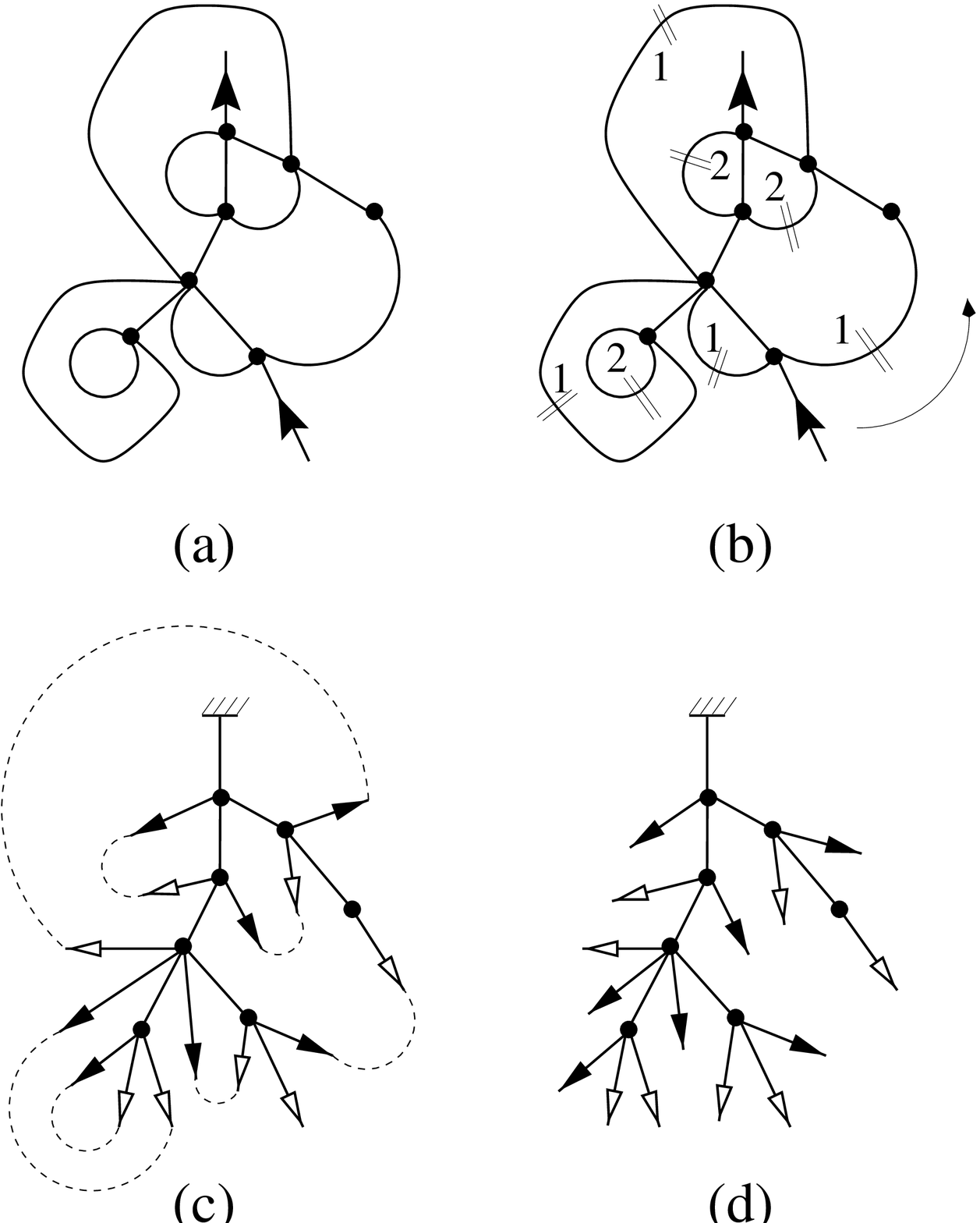}{10.cm} 
\figlabel\oneone

The one-to-one correspondence is illustrated in Fig.\oneone.
To each two-leg diagram we associate a rooted blossom tree as follows.
Starting from the in-coming leg, we iteratively visit all edges of the diagram
adjacent to the external face in {\it counterclockwise} direction. At each
step we cut the edge iff the remaining diagram remains connected. The first half
of the edge is replaced by a bud, the second one by a leaf. 
This procedure is repeated with the remaining diagram until a tree is obtained.
Finally, the in-coming leg is replaced by a leaf, while the out-coming one 
is chosen to be the  root. It is shown in Refs.[\xref\SCH,\xref\CENSUS] 
that the resulting tree is indeed a blossom tree as defined above.
This transformation is moreover invertible, by iteratively connecting 
each bud to the closest available leaf in counterclockwise direction
and fusing these bud-leaf pairs into edges. 
This construction leaves one unmatched leaf, adjacent to the external face, and
that we replace by the in-coming leg, while the root becomes the out-coming one. 
Note that in the above fusion procedure the
root may possibly be encompassed by a number of nested edges,
in which case it does not lie in the external face. 
The number of ``encompassing" edges will be called the {\it depth}
of the root in the blossom tree.

The generating function $R$ for blossom trees with a weight $g_k$ per $2k$-valent
vertex satisfies
\eqn\satisR{ R=1+\sum_{k\geq 1} g_k  {2k-1\choose k-1} R^k }
as obtained by inspecting all possible configurations of the vertex attached to the root.
The function $R$ is uniquely determined as the formal power series solution
of eq.\satisR\ in the $g_k$'s with constant term $1$. 
By the above correspondence, $R$ also coincides with the generating function
$\Gamma_{1,1}$ of two-leg diagrams with a weight $g_k$ per inner $2k$-valent vertex.
This is to be contrasted with the generating function $\Gamma_2$ for
two-leg diagrams with both legs in the external face, or equivalently
blossom trees with root at depth $0$, which reads \CENSUS\
\eqn\gammatwo{ \Gamma_2 = R -\sum_{k\geq 2} g_k {2k-1\choose k-2} R^{k+1} }

\newsec{Geodesic distance}

A nice feature of the above bijection is that it keeps track 
of the {\it geodesic distance} between the in- and out-coming legs
in the two-leg diagrams, defined as the minimal number of edges 
crossed by a curve connecting the two legs (for instance, the two legs of the diagram
of Fig.\oneone\ (a) are at distance $1$). Indeed, this distance 
is nothing but the depth of the root in the corresponding blossom trees.
More precisely, it was shown in Refs.[\xref\SCH-\xref\BMS] 
that all the edges cut in the cutting algorithm belong to minimal
paths (i.e. paths crossing a minimal number of edges)
going from the external face to all internal ones. In particular, the
path from the external face (adjacent to the incoming leg) to that adjacent 
to the out-coming leg and crossing each 
encompassing edge exactly once is minimal. 

This leads to the refined definition of the generating function $G_n$ for
blossom trees with root at depth $n$, which alternatively
generates the two-leg diagrams with geodesic distance $n$ between the legs. 
In particular, we have $\Gamma_2=G_0$ and $\Gamma_{1,1}=\sum_{n\geq 0} G_n$.
In the following it will prove easier to use the generating 
function $R_n$ for blossom trees with root at depth {\it at most} $n$,
such that 
\eqn\wehaR{\eqalign{ 
&G_n= R_n -R_{n-1} \cr
&\Gamma_2= R_0 \cr
&\Gamma_{1,1}=R=\lim_{n\to \infty} R_n \cr}}
with the convention that $R_{-1}=0$.

The first result of this paper is a refined version of eq.\satisR\ 
which keeps track of the depth of the root of the trees.

\fig{The contour walk (b) associated to a blossom tree (a) obtained by visiting the buds
and leaves around the tree in clockwise direction starting from the root. Each bud (resp. leaf) 
corresponds to a $+1$ (resp. $-1$) step. The walk is decomposed in (c) according
to the subtrees attached to the root vertex, here two buds and
three blossom trees of respective root depths $0$, $2$, $0$, viewed as the
maximal relative heights reached by the corresponding portions of walk.
Here, the depth of the root is $3$, which corresponds to the maximal
height attained by the contour walk.}{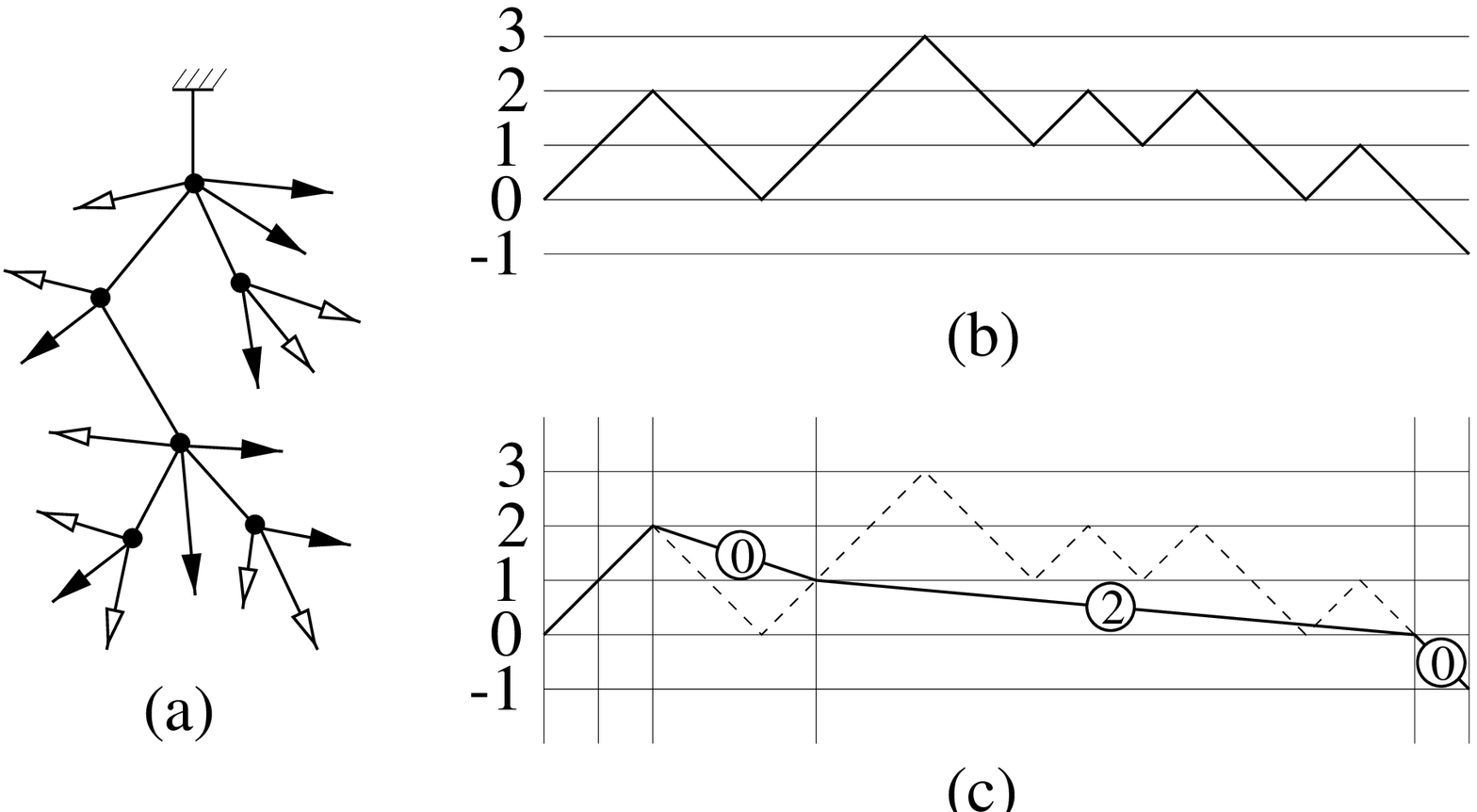}{13.cm}
\figlabel\Rwalk

For any blossom tree, let us visit all its buds and leaves successively in clockwise
order around the external face starting from the root.
This defines the {\it contour walk} of the tree, namely a walk on the integer line 
starting from $0$ and moving up (height shifted by $+1$)
for each bud encountered and down (height shifted by $-1$) for each leaf.
As leaves are in excess of one on each of these trees, the contour walk stops at height $-1$.   
We now easily identify the depth of the root as the maximal height 
reached by the walk (see Fig.\Rwalk).

\fig{The root vertex environment is characterized by the cyclic sequences of attached buds
and blossom subtrees, represented with a circle. The index $m_i$ denotes the number of buds 
immediately before the $i$-th blossom subtree, whose root depth is denoted by $p_i$.
The corresponding countour walk is also sketched, and gives a visual representation of eq.(3.2).}
{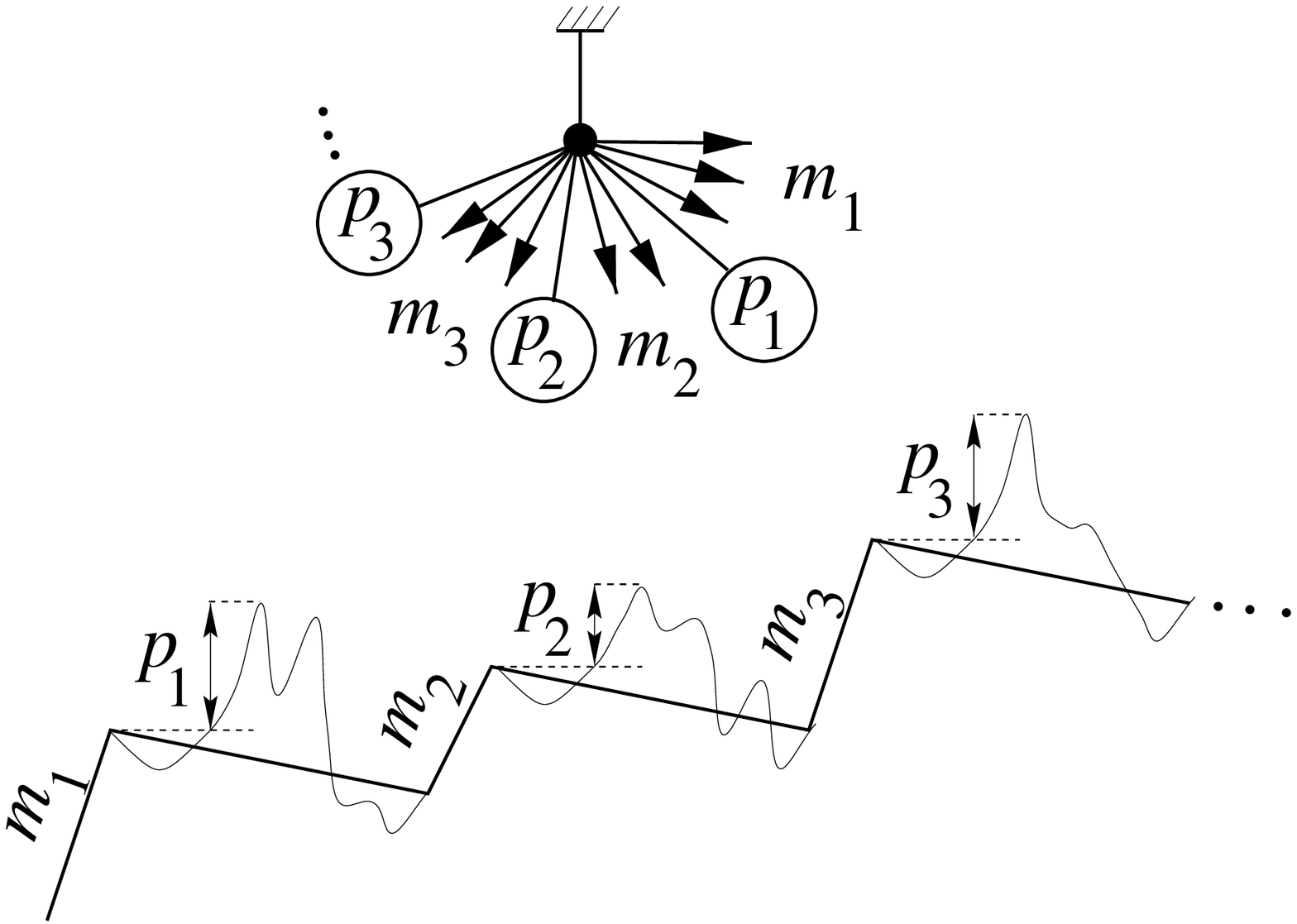}{12.cm}
\figlabel\environ

As before, a recursion relation is obtained by inspection of all possible environments
of the root in a given blossom tree. In the case of the smallest blossom tree made of 
a single leaf, the depth of the root is $0$, hence this tree contributes $1$ to
$R_n$ for all $n\geq 0$.
In the case of a tree with a $2k$-valent
vertex attached to the root, we have exactly $k-1$ buds and $k$ descendent blossom trees 
attached (see Fig.\environ).
Turning around the vertex in clockwise direction from the root, 
let $m_i$ denote the (possibly vanishing) number of consecutive buds
immediately before the $i$-th descendent blossom tree, 
$i=1,2,...,k$ (with necessarily
$\sum m_i \leq k-1$, $m_i\geq 0$). Let moreover $p_i$, $i=1,2,...,k$ denote the
depth of the root of the $i$-th descendent blossom tree,
considered as an independent rooted blossom tree. With these definitions,
the depth $p$ of the root of the whole tree reads
\eqn\eblp{ p=\max_{1\leq i\leq k}\left\{ p_i+m_i+m_{i-1}+\cdots+m_1-(i-1)
\right\}}  
This relation is clear in the contour walk picture, where 
the walk associated to the whole tree decomposes into $+1$ steps for the buds attached
to the vertex, and into subwalks associated with the descendent blossom trees, each 
contributing by a global $-1$ height shift. The quantity $p_i+m_i+m_{i-1}+\cdots+m_1-(i-1)$
is nothing but the maximal height reached within the $i$-th subwalk, which 
starts from height $m_i+m_{i-1}+\cdots+m_1-(i-1)$.    
From eq.\eblp, we see that the condition $p\leq n$
is equivalent to $p_i \leq n+i-1-m_1-m_2-\cdots-m_i$ for all $i=1,2,...,k$, which finally
translates into the recursion relation
\eqn\recurela{ R_n = 1 +\sum_{k\geq 1} \ \  g_k \! \! \! \sum_{m_i\geq 0\atop
m_1+...+m_k\leq k-1} \prod_{i=1}^k R_{n+i-1-m_1-\cdots-m_i}  }
with the convention $R_n=0$ for $n<0$.
This may be viewed as a generalization of eq.\satisR\ which is recovered in the limit 
$n\to \infty$, where $R_n\to R$.

The equation \recurela\ may be rewritten in an algebraic way by use of a shift
operator $\sigma$ acting on the canonical basis $\{p_n\}_{n\in \IZ}$ for 
sequences as $\sigma p_n=p_{n+1}$, and of its
inverse $\sigma^{-1}$ such that  $\sigma^{-1}p_n=p_{n-1}$, namely introducing the operator
\eqn\opq{ Q= \sigma +\sigma^{-1} r }
where $r$ acts diagonally as $r p_n=R_n p_n$, eq.\recurela\ takes the form 
\eqn\recumieux{ Q_{n-1,n}=1 + \sum_{k\geq 1} g_k (Q^{2k-1})_{n-1,n} }
where $Q_{i,j}$ stand for the matrix elements in the canonical basis $\{p_n\}$.
This algebraic formulation \recumieux\ is very similar to the main recursion
relation obtained in the context of one-matrix models with ${\cal N}\times {\cal N}$
Hermitian matrices and with even potentials of the
form $V(M)={M^2\over 2} -\sum_{k\geq 1} g_k {M^{2k}\over 2k}$, known to generate
fatgraphs with $2k$-valent vertices weighted by $g_k$ \DGZ. In this
setting, the operator $Q$ represents the multiplication by an eigenvalue $\lambda$,
acting on polynomials of $\lambda$. In the basis made of the monic 
polynomials $p_n(\lambda)$ orthogonal \wrt the measure
$d\lambda e^{-{\cal N} V(\lambda)}$ on $\IR$, $Q$ takes the form \opq\ and the recursion 
relation for the $R_n$'s takes precisely the form \recumieux, except that the first term 
on the rhs is changed from 
$1$ to $n/{\cal N}$. This coincidence is both remarkable and mysterious as in the matrix model
recursion, the $R_n$'s  give access to the generating functions for higher genus graphs, as opposed
to the present ones, which only concern planar graphs, but on the other hand allow to explore 
geodesic distances, a task out of reach in the matrix model.

\newsec{Tetravalent case}

\subsec{Exact solution}

Let us now show how eq.\recurela\ may be used to extract exact expressions
for the $R_n$'s. As a first simpler but instructive case, let us restrict
our study in this section to the case of {\it tetravalent} maps, i.e.
two-leg diagrams having only tetravalent inner vertices. Going to 
generating functions, this amounts to taking $g_k=g \delta_{k,2}$, in which
case eq.\satisR\ reduces to:
\eqn\satisRtetra{R=1+3gR^2}
with solution
\eqn\Rtetra{R={1-\sqrt{1-12g}\over 6g}}
which displays the well-known critical value $g_c=1/12$ of $g$ for pure 
quadrangulations. 
More generally, the relation \recurela\ reduces to:
\eqn\recutetra{R_n=1+gR_n(R_{n-1}+R_n+R_{n+1})}
Conceptually, this equation, valid for all $n\geq 0$, 
may be viewed as a recursion of order
$2$ in $n$, fully determining all the $R_n$'s from the initial
conditions:
\item{(i)} $R_{-1}=0$;
\item{(ii)} $R_0=\Gamma_2$, where $\Gamma_2$ is obtained from
eqs.\gammatwo\ and \satisR, namely $\Gamma_2=R-gR^3$ with $R$
as in \Rtetra.
\par 
\noindent From the combinatorial origin of the equation, we expect that 
the non trivial initial condition (ii)  may be replaced by the 
alternative boundary condition at infinity that:
\item{(ii')} $\lim_{n\to\infty} R_n=R$ with $R$ as in \Rtetra.
\par
The general solution to eq.\recutetra\ satisfying (ii') reads:
\eqn\solquart{R_n=R{(1-\lambda x^{n+1})(1-\lambda x^{n+4})
\over (1-\lambda x^{n+2})(1-\lambda x^{n+3})}}
where $\lambda$ is an arbitrary ``integration" constant and 
$x$ is the solution of the characteristic equation:
\eqn\linqua{ {1-4gR \over gR} =x+{1\over x} }
with $|x|<1$. Note that for $0\leq g\leq g_c$, the lhs is
larger or equal to $2$, hence $x$ is real and positive. 
The condition $0\leq x<1$ (which ensures the convergence (ii')
of $R_n$ to $R$) can always be satisfied for $g<g_c$. 

That the expression
\solquart\ solves eq.\recutetra\ may be checked explicitly.
In a more constructive way, this solution is obtained 
by expanding $R_n$ in the vicinity of $R$ as $R_n=R(1-\rho_n)$   
and first expressing eq.\recutetra\ at first order in $\rho_n$
with the resulting linearized equation
\eqn\linear{\rho_n^{(1)}=gR(\rho_{n-1}^{(1)}+\rho_n^{(1)}+
\rho_{n+1}^{(1)})+3gR\rho_n^{(1)}}
The general solution of this linear recursion such that 
$\lim_{n\to\infty}\rho_n^{(1)}=0$ has the form $\rho_n^{(1)}=\alpha_1 x^n$
where $x$ is the solution of the characteristic equation \linqua\ with
$|x|<1$. Further expanding $R_n$, we have
\eqn\exparho{\rho_n=\sum_{i\geq 1}\alpha_i (x^n)^i}
Substituting $R_n=R(1-\rho_n)$ and picking the term of order $k+1$ 
in $x^n$ yields the relation
\eqn\recrho{\alpha_{k+1}\left(x^{k+1}+{1\over x^{k+1}}-x-{1\over x}\right)
=\sum_{j=1}^k \alpha_j \alpha_{k+1-j}\left(x^j+{1\over x^j}+1\right)}
solved recursively as 
\eqn\solarho{\alpha_k=\alpha_1\left({\alpha_1 x\over (1-x)(1-x^2)}\right)^{k-1}
{1-x^k\over 1-x}}
Picking $\alpha_1=x(1-x)(1-x^2)\lambda$, $R_n$ is easily resummed
into \solquart.

Further imposing the initial condition (i) fixes $\lambda=1$, hence
\eqn\soltet{\eqalign{R_n&=R{(1-x^{n+1})(1-x^{n+4})
\over (1-x^{n+2})(1-x^{n+3})} \cr
&= R {U_n(w)U_{n+3}(w)\over U_{n+1}(w)U_{n+2}(w)}\cr}}
where the $U_n$'s are the Chebyshev polynomials of the first kind,
namely obeying
$U_{n+1}(w)=w U_n(w)-U_{n-1}(w)$ and $U_0=1$, $U_1=w$,
expressed in the variable 
\eqn\aequal{ w=\sqrt{x}+{1\over \sqrt{x}} =\sqrt{1-2gR\over gR}}

A first outcome of this solution is that it allows to recover 
the value of $\Gamma_2$, namely
\eqn\initqua{\Gamma_2=R_0 =R {x+{1\over x} \over 1+x+{1\over x}}=R {1-4gR\over 1-3gR}
=R-gR^3}
as expected, 
which {\it a posteriori} shows the equivalence (ii) $\leftrightarrow$ (ii'). 

\fig{The tetravalent two-leg diagrams with two inner vertices, classified according
to the geodesic distance $n$ between the two legs, namely $n=0$ (a), $n=1$ (b)
and $n=2$ (c).}{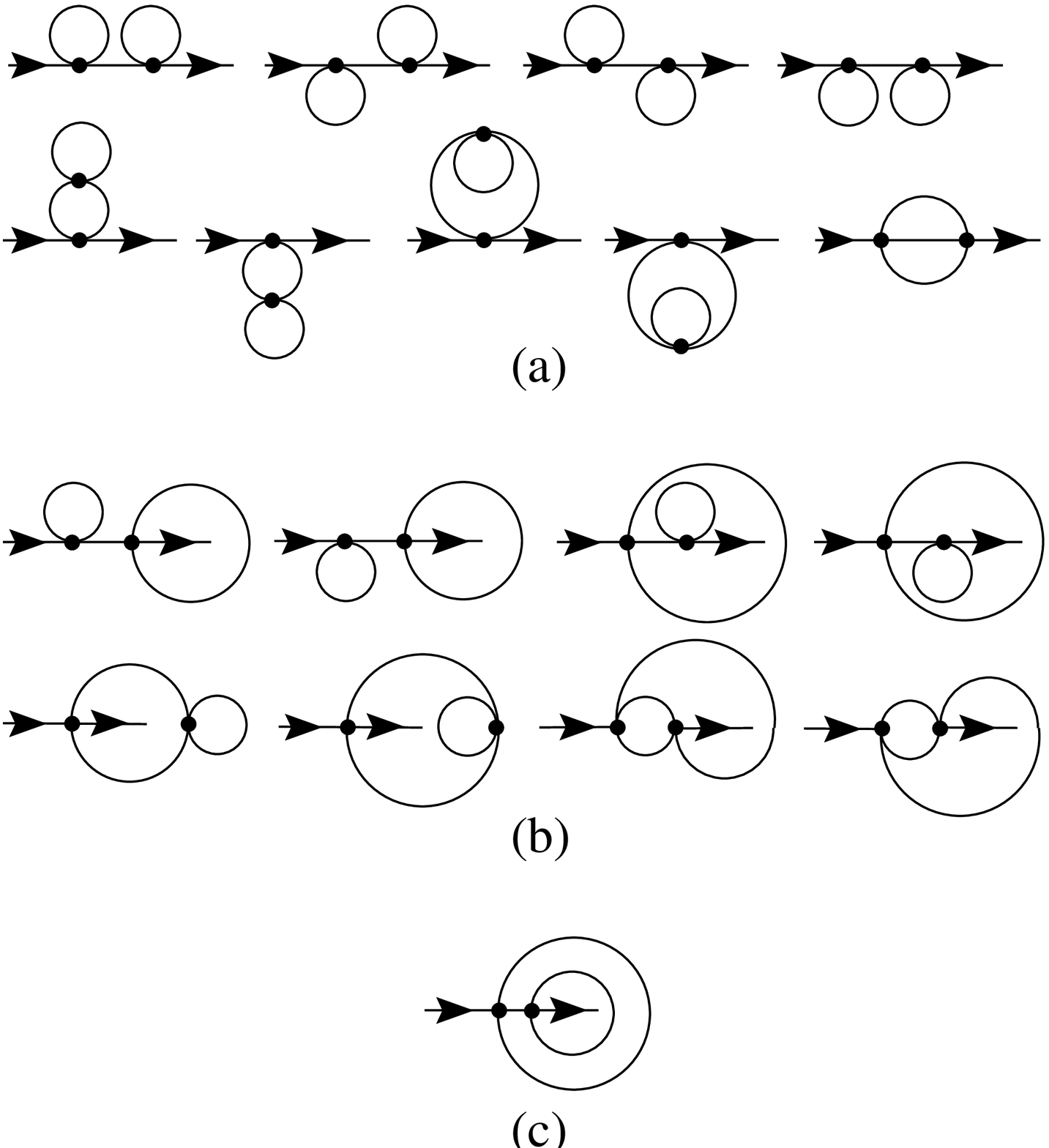}{10.cm}
\figlabel\firstfew

As an illustration of the result \soltet, we list below the first few generating
functions $G_n(g)=R_n-R_{n-1}$ for two-leg diagrams with legs at a
geodesic distance $n=0,1,2$... 
\eqn\firfew{\eqalign{
G_0(g)&={24g-1+\sqrt{1-12g} \over 9 g(1+\sqrt{1-12g})}=1+2g+9g^2+54 g^3
+378 g^4+2916 g^5+\cdots\cr
G_1(g)&={27g-2+(2-15g)\sqrt{1-12g}\over 27 g(1-8g+\sqrt{1-12g})}=
g+8g^2+65g^3+554g^4+4922g^5+\cdots\cr
G_2(g)&={252 g^2-6g-1+(1+12g)\sqrt{1-12g}\over 27(190g^2-51g+3+(3-33g+44g^2)\sqrt{1-12g})}\cr
&=g^2+15g^3+179g^4+1995g^5+\cdots\cr}}
We have represented in Fig.\firstfew\ all the two-leg diagrams with 2 vertices,
among which 9 (resp. 8, 1) have the two legs at geodesic distance 0 (resp. 1, 2),
in agreement with the $g^2$ terms in eq.\firfew.

\subsec{Fractal dimension}

As a direct application of the solution \soltet, let us evaluate the
``fractal dimension" of large tetravalent planar maps, defined via the number 
$R_{n,N}$ of two-leg diagrams
with distance less or equal to $n$ 
between the two legs and with $N$ vertices.
Indeed, at large $N$, the limiting ratio 
\eqn\behanum{ B_n\equiv \lim_{N\to \infty} {R_{n,N}\over R_{0,N}} } 
may be taken as a good estimate of the average number of points 
at a geodesic distance less or equal to $n$ from
a given point in random tetravalent maps of infinite size. 
It is expected to behave like 
\eqn\behas{ B_n \sim n^{d_F} \ \ {\rm for} \ {\rm large}\ n }
where $d_F$ is the fractal dimension.

We have 
\eqn\finrn{
R_{n,N}= \oint {dg \over 2i\pi g^{N+1}} R_n }
where we must use the expression \soltet\ for $R_n$, in which we substitute expression
\Rtetra\ for $R(g)$ and solve eq.\linqua\ for 
$x$ as a function of $g$.
Upon performing the change of variable $g\to v=gR(g)$, namely $g(v)=v(1-3v)$,
we arrive at  
\eqn\varchange{ R_{n,N}= \oint {dv (1-6v) \over 2i\pi (v(1-3v))^{N+1}} 
{1\over 1-3v} {(1-x(v)^{n+1}) (1-x(v)^{n+4}) \over 
(1-x(v)^{n+2})(1-x(v)^{n+3})} } 
where we have used $R(g(v))=1/(1-3v)$ and the expression 
$x=x(v)\equiv (1-4v-\sqrt{1-8v+12 v^2})/(2v)$.
The large $N$ behavior is obtained by
a saddle-point approximation, with the result
\eqn\RnN{ R_{n,N} \sim {\rm const.}{(12)^N \over N^{5\over 2}} 
{(n+1)(n+4)\over (n+2)(n+3)} (140+270 n+179 n^2+50 n^3+5 n^4) }
which finally gives the ratio 
\eqn\bnexpr{ B_n = {3\over 280} {(n+1)(n+4)\over (n+2)(n+3)} 
(140+270 n+179 n^2+50 n^3+5 n^4)
\sim {3\over 56} n^4 }
which yields $d_F=4$, as expected for pure gravity.

\subsec{Critical scaling}

A continuum limit may be reached by letting $g$ tend
to its critical value $g_c=1/12$,
corresponding to  $gR\to 1/6$ and $x\to 1$. 
More precisely, we write
\eqn\wriscaqua{ g={1\over 12}(1-\epsilon^4) \qquad gR={1\over 6}(1-\epsilon^2)}
{}from eq.\Rtetra. In turn, the characteristic equation \linqua\ yields
\eqn\xrt{ x=e^{-a\epsilon}+O(\epsilon^2) \qquad a=\sqrt{6} }
As seen from eq.\soltet, a sensible limit is obtained by writing
\eqn\scaling{n={r\over \epsilon}}
and letting $\epsilon\to 0$. Writing the scaling variable $r$ as $r=n/\xi$,
we see that $\epsilon$ plays the role of the inverse of the
correlation length $\xi$. 
As we approach the critical point, we have 
$\xi=\epsilon^{-1}= \big((g_c-g)/g_c\big)^{-\nu}$ with a critical exponent  
$\nu=1/4$, in
agreement with $\nu=1/d_F$, as expected from general principles. 
Performing this limit explicitly on the solution \soltet\ yields
an explicit formula for the continuum partition function ${\cal F}(r)$ of
surfaces with two marked points
at a geodesic distance {\it larger or equal to} $r$:
\eqn\exprneps{ {\cal F}(r)\equiv \lim_{\epsilon\to 0}{R-R_n\over \epsilon^2 R}=
-2 {d^2 \over dr^2} {\rm Log}\, \sinh \left({\scriptstyle\sqrt{3\over 2}}\, r\right)={3
\over \sinh^2\left({\scriptstyle\sqrt{3\over 2}}\, r\right)} }
Upon differentiating \wrt $r$, we obtain the continuum partition function
for surfaces with two marked points at a
geodesic distance {\it equal to} $r$: 
\eqn\exprderiv{{\cal G}(r)=-{\cal F}'(r)=
{3\sqrt{6}}\, {\cosh \left({\scriptstyle\sqrt{3\over 2}}\, r\right)\over \sinh^3
\left({\scriptstyle\sqrt{3\over 2}}\, r\right)}}
in agreement with the form obtained in Ref.\AW.

Note that the precise form of the scaling function ${\cal F}(r)$ may alternatively
be obtained by solving the continuum counterpart of eq.\recutetra. Indeed, 
writing 
\eqn\scaleq{R_n=R(1-\epsilon^2 {\cal F}(n\epsilon))}
and expanding eq.\recutetra\ up to order $4$ in $\epsilon$, we obtain the
following differential equation
\eqn\painleve{{\cal F}''(r)-3{\cal F}^2(r)-6{\cal F}(r)=0}
It is easy to check that ${\cal F}(r)$ as given by \exprneps\ is the 
unique solution of \painleve\ with boundary conditions ${\cal F}(r)\to \infty$
when $r\to 0$ and ${\cal F}(r)\to 0$ when $r\to \infty$.

\subsec{Continuum probability distribution for geodesic distances}

The scaling function ${\cal G}(r)$ obtained in the previous subsection is the continuum 
``two-point function" for surfaces with two marked points at geodesic distance $r$.
It is obtained in a grand-canonical formalism, in which the area of the surfaces
(number of vertices in the corresponding maps) is not fixed.   
In particular, ${\cal G}(r)$ cannot be interpreted as a probability distribution
as it is not normalizable. Such a problem does not occur in the  
{\it canonical} formalism, i.e. when working with maps of fixed area $N$. 
A well defined probabilty is given by
$R_{n,N}/R_{\infty,N}$,  where $R_{n,N}$ is the number of two-leg diagrams 
with $N$ inner vertices and legs distant by at most $n$, as before,
and $R_{\infty,N}$ is the {\it total} number of two-leg diagrams with 
$N$ inner vertices.
Starting from the expression \finrn\ for $R_{n,N}$ and performing
again the change of variables $g=v(1-3v)$, we get
\eqn\getch{ 
R_{n,N}= \oint {dv (1-6v) \over 2i\pi (v(1-3v))^{N+1}} R_n(v) }
where the integral is to be taken on a small contour around the origin.
At large $N$, the integral is dominated by the saddle-point $v=V_c=1/6$.
We therefore perform the change of variables $v=V_c(1+iu/\sqrt{N})$, 
with $u$ to be integrated on the real line when $N\to \infty$.
As opposed to the computation of Sect.4.2 where $R_{n,N}$ was 
estimated for {\it finite} $n$ and large $N$, we may obtain a
non-trivial distribution by letting $N \to \infty$ while $n$ scales
as $n=\alpha N^{1/4}$, in which case we may use eq.\scaleq\ with 
$\epsilon=\sqrt{-i u}/N^{1/4}$, namely
\eqn\rewrn{\eqalign{
R_n(v)&=R\left(1+i{u\over \sqrt{N}} {\cal F}(\alpha \sqrt{-iu})+O({1\over N})\right)\cr
&={V_c\over g_c}\left(1+{iu\over \sqrt{N}}+O({1\over N})\right)\left(1+i{u\over \sqrt{N}} {\cal F}(\alpha
\sqrt{-iu})+O({1\over N})\right)\cr
&=2\left(1+{iu\over \sqrt{N}}\left(1+{\cal F}(\alpha \sqrt{-iu})\right)\right)+O({1\over N})\cr}}
Substituting this into eq.\getch, we find
\eqn\findg{\eqalign{
R_{n,N}&\sim 2{12^{N}\over \pi N^{3/2}}\int_{-\infty}^\infty du 
u^2 e^{-u^2}\left(1+{\cal F}(\alpha \sqrt{-iu})\right)\cr
&=4{12^{N}\over \pi N^{3/2}}\int_{0}^\infty du
u^2 e^{-u^2}\left(1+{\rm Re}\left({\cal F}(\alpha \sqrt{-iu})\right)\right)\cr
}}   
and a similar expression for the limit $R_{\infty,N}$, obtained by taking ${\cal F}\to 0$
in the above. This leads to the probability $\Phi(\alpha)$ for 
a random two-leg diagram to have legs at a rescaled
geodesic distance less than $\alpha$:
\eqn\probaqua{ \Phi(\alpha)={2\over \sqrt{\pi}} \int_0^\infty du u^2 e^{-u^2}
{\cosh(2\alpha\sqrt{3u})+\cos(2\alpha\sqrt{3u})+8 \cosh(\alpha\sqrt{3u})\cos(\alpha\sqrt{3u})
-10 \over \left(\cosh(\alpha\sqrt{3u})-\cos(\alpha\sqrt{3u})\right)^2} }
Some remarks are in order concerning this explicit form. First
we may explicitly expand the probability distribution $\Phi(\alpha)$ as a
formal power series of $\alpha$, with the result
\eqn\Phial{ \Phi(\alpha)=6\sum_{j=1}^\infty (-1)^{j+1} {\alpha^{4j} 3^{2j}} 
{(2j)!\over j! (4j)!} B_{4j+2}  }
where $B_j$ denote the Bernoulli numbers. 
This series has a radius of convergence zero but provides an 
asymptotic expansion valid to any finite order.
In particular, the leading small $\alpha$ behavior is $\Phi(\alpha)\sim 3 \alpha^4/28$,
which translates into the equivalent $R_{n,N}\sim (3/28)(n^4/N) R_{\infty,N}$ for
finite $n$ and large $N$, while from eq.\bnexpr\ we have $R_{n,N}\sim (3/56)n^4 R_{0,N}$.    
These relations are compatible, as we know that $R_{0,N}=2 R_{\infty,N}/(N+2)$ from
combinatorial arguments \SCH. 

Secondly, upon differentiating with respect to $\alpha$, we obtain the probability 
distribution $\rho(\alpha)$ for two-leg diagrams with legs at rescaled
geodesic distance $\alpha$, with the result:
\eqn\probadifqua{\eqalign{\rho(\alpha)={4\over \sqrt{\pi}\alpha} \int_0^\infty &du u^2
(2u^2-3) e^{-u^2} \times \cr &\times
{\cosh(2\alpha\sqrt{3u})+\cos(2\alpha\sqrt{3u})+8 \cosh(\alpha\sqrt{3u})\cos(\alpha\sqrt{3u})
-10 \over \left(\cosh(\alpha\sqrt{3u})-\cos(\alpha\sqrt{3u})\right)^2}\cr} }
At large $\alpha$, the function $\rho(\alpha)$ 
decays as $\exp(-C \alpha^{\delta})$, with $C=3(3/8)^{2/3}$ and
with a critical exponent $\delta=4/3$, in agreement with Fisher's law $\delta=1/(1-\nu)$, 
a result already observed in Ref.\AW.

\fig{Plots of the probability distributions $\Phi(\alpha)$ (a) and $\rho(\alpha)$ (b) 
for two-leg diagrams with legs at rescaled geodesic distance at most (a) or equal to (b)
$\alpha$.}{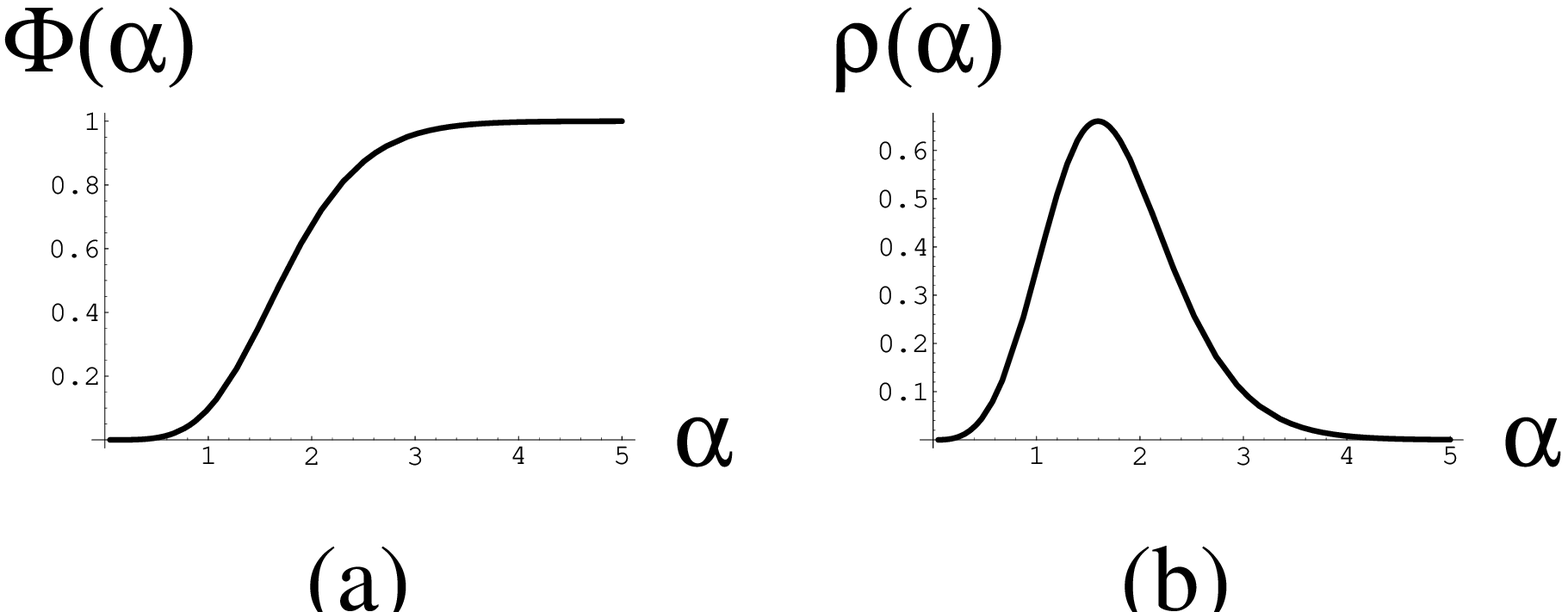}{13.5cm}
\figlabel\phirho
\noindent The distributions $\Phi(\alpha)$ and $\rho(\alpha)$ are plotted
in Fig.\phirho\ for illustration.

\newsec{Exact solution in the general case}

Let us now turn to the general case of maps with bounded valences,
namely with arbitrary even valences $2,4,6,...$ up to say $2(m+1)$.
This corresponds to keeping arbitrary vertex weights $g_1,g_2,...,g_{m+1}$
while setting $g_j=0$ for $j>m+1$. 
The equation \recurela\ becomes a recursion relation of order $2m$
namely expressing $R_{n+m}$ in terms of $R_{n+j}$, $j=m-1,m-2,...,-m$.
This equation is valid for all $n\geq 0$, 
upon taking the initial condition
\item{(i)} $R_{-1}=R_{-2}=...=R_{-m}=0$.
\par
\noindent To completely fix the solution, the condition (i) must be supplemented by 
\item{(ii)} the data of the $m$ first terms $R_0,R_1,...,R_{m-1}$.
\par
\noindent The explicit combinatorial expression for $R_0$ is known from the identification
with the two-leg function $\Gamma_2$, given by eq.\gammatwo. As to the other $R_j$,
$j=1,2,...,m-1$, no direct combinatorial expression is available yet.
In view of the previous section, and from the combinatorial interpretation
of $R_n$, we replace the initial data (ii) above with
the condition
\item{(ii')} $\lim_{n\to \infty} R_n =R$, where $R$ is the solution of eq.\satisR.
\par
\noindent As we shall see, this latter condition is indeed sufficient, 
together with initial condition (i) to completely fix the solution $R_n$. 

To reach this solution, we follow the same strategy as in the tetravalent case of Sect.4.
We first obtain constructively the solutions to eq.\recurela\ with the only boundary
condition (ii'), expressed in terms of $m$ free
``integration constants" $\lambda_1,...,\lambda_m$. We finally implement the boundary 
condition (i) to further fix these constants.

Starting with the condition (ii'), let us first expand $R_n$ around its limit
as $R_n=R(1-\rho_n)$ and express eq.\recurela\ at first order in $\rho_n$
\eqn\lineariz{\eqalign{ \rho_n^{(1)}&=\sum_{k=0}^{m} g_{k+1} R^k \sum_{j=-k}^k
\left(\sum_{l=|j|\atop l=j {\rm mod}\ 2}^{2k-|j|} 
{l\choose {j+l\over 2}} {2k-l\choose {j+2k-l\over 2}}
\right)\rho_{n+j}^{(1)}\cr  
&=\sum_{k=0}^{m} g_{k+1} R^k \sum_{j=-k}^k  
\left(\sum_{l=0}^{k-|j|} {2k+1 \choose l}  
\right)\rho_{n+j}^{(1)}\cr}}
The coefficient of $\rho_{n+j}^{(1)}$ in the first line is easily obtained in the contour walk
picture as the number of walks from height $n$ to height $n-1$ with $2k+1$ steps
and a marked step down from height $n+j$, itself decomposed into 
a walk from height $n$ to $n+j$ and say $l$ steps, and a walk from height $n+j-1$
to $n-1$ with $2k-l$ steps.
Looking for a solution $\rho_{n+j}^{(1)}=\alpha_1 x^n$, we get the characteristic equation
\eqn\chargen{\eqalign{ \chi_m(x)=0,\qquad \chi_m(x)&\equiv 1 -\sum_{k=0}^m  g_{k+1} R^{k} 
\sum_{l=0}^{k} {2k+1\choose l} {1\over x^{k-l}} {1-x^{2k+1-2l}\over 1-x}\cr
&=1 -\sum_{k=0}^m  g_{k+1} R^{k} \sum_{l=0}^{k} {2k+1\choose l} U_{2k-2l}(w) \cr}}
where the $U$'s are the Chebyshev polynomials of the first kind,  
with again $w=\sqrt{x}+1/\sqrt{x}$.
Note that $\chi_m(x)$ is invariant under the interchange $x\leftrightarrow 1/x$,
and moreover is a polynomial of degree $m$ in $x+{1/x}$. The most general
solution $\rho_{n}^{(1)}$ is a linear combination
\eqn\lincom{ \rho_{n}^{(1)}=\sum_{j=1}^m \alpha_j x_j^n }
where $x_j$, $j=1,2,...,m$ are the $m$ zeros of $\chi_m$ with modulus less than $1$.
Further expanding $\rho_n$ order by order in the $x_j^n$, as
$\rho_n=\sum_{n_1,...,n_m\geq 0} \alpha_{n_1,...,n_m}  (x_1^n)^{n_1} ... (x_m^n)^{n_m}$
with $\alpha_{0,0,...,0}=0$ and $\alpha_{0,..,0,1,0,...,0}=\alpha_j$ 
for the $1$ in position $j$,
we get a recursion relation for the coefficients $\alpha_{n_1,...,n_m}$, which determines
them all in terms of the initial integration constants $\alpha_1,...,\alpha_m$ of \lincom.
Upon redefining them like in previous section as
$\alpha_i=x_i(1-x_i)(1-x_i^2) \lambda_i$, and resumming the expression for $R_n$,
we arrive at
\eqn\mxsol{\eqalign{
&R_n=R{u_n^{(m)} u_{n+3}^{(m)} \over u_{n+1}^{(m)} u_{n+2}^{(m)} } \cr
&u_n^{(m)}= \sum_{l=0}^m (-1)^l \sum_{1\leq m_1 <...<m_l\leq m}
\prod_{i=1}^l \lambda_{m_i} x_{m_i}^{n+m} \prod_{1\leq i<j\leq l} c_{m_i,m_j}\cr
&c_{a,b}\equiv {(x_{a} -x_{b})^2 \over (1-x_{a} x_{b})^2} \cr
}} 
where, as before, the $x_i$ must be chosen so that
$\chi_m(x_i)=0$, $|x_i|<1$, $i=1,2,...,m$. 
Proving this result is beyond the scope of the present paper. That the form \mxsol\
actually solves \recurela\ with $x$'s solutions of eq.\chargen\ with modulus less
than 1 may be explicitly checked for small values of $m=2,3,4$. 

A few remarks are in order regarding the particular form \mxsol. 
First the expression \mxsol\ as a function of $R$ and of the $x_i$'s
is completely independent of the particular values of the parameters $g_k$. 
The link with the specific eq.\recurela\ appears only through the expression
for $R$ solving \satisR\ and the characteristic equation \chargen\ 
satisfied by the $x$'s.
The form of $u_n^{(m)}$ is reminiscent of that of the so-called
N-soliton tau-function of the KP hierarchy \JM\ with the generic form
\eqn\jimiw{ \eqalign{
&\tau=\sum_{r=0}^N \sum_{i_1<...<i_r} \prod_{\mu=1}^r e^{\eta_{i_\mu}}
\left(\prod_{\mu<\nu} c_{i_\mu,i_\nu}\right) \cr
&c_{i,j}={(p_i-p_j)(q_i-q_j)\over (p_i-q_j)(q_i-p_j)}\cr}}
where the $\eta$'s contain the times' dependence of the KP hierarchy. 
We see that, at least from a formal point of view, our solution \mxsol\ 
corresponds to taking $N=m$, $e^{\eta_i}=-x_i^{n+m} \lambda_i$, $p_i=x_i$
and $q_i=1/x_i$.
This relation, though suggestive, remains mysterious at present.   
Secondly, we note that any ``N-soliton" solution of the form \mxsol\ 
with $m\to N>m$ reduces to the
$m$-soliton solution when we pick the $x$'s among the $m$ zeros of $\chi_m$ with
modulus less than 1. Indeed, for $N>m$,
some of the $x$'s will be equal resulting in the vanishing of the
corresponding $c_{i,j}$'s.
Analogously, taking an $N<m$ solution is perfectly licit, as it amounts to 
taking some $\lambda$'s to zero, but it corresponds to a less generic
solution. In particular, the one-soliton solution \solquart\
is solution to all general equations, provided $x$ solves the corresponding 
characteristic equation.
Thirdly, inspired by the determinantal form of various soliton expressions, we find
that $u_n^{(m)}$ of eq.\mxsol\ may be written as the following determinant 
\eqn\detergent{\eqalign{ &u_n^{(m)}=\det\  M_n^{(m)} \cr
&\big[M_n^{(m)}\big]_{i,j}=\delta_{i,j} -\lambda_i x_i^{n+m} 
{1-x_i^2\over 1-x_ix_j}\qquad i,j=1,2,...,m \cr} }
as proved in Appendix A below.

Eq.\mxsol\
provides the general solution to \recurela\ {\it a priori}
valid for any $n$ (not necessarily positive or zero), 
that satisfies the convergence condition (ii').
Further imposing the initial conditions (i) also implies, by writing \recurela\
for $n=-1,-2,...-(m-1)$ that $R_{-m-1}$ and $R_{-m-2}$ diverge and that
$R_{-m-3},R_{-m-4},...,R_{-(2m-1)}$ vanish,
which in turn entail that $u_{-1}^{(m)}=0$, $u_{-2}^{(m)}=0$, ...$u_{-(2m-1)}^{(m)}=0$.
These latter conditions entirely 
fix the values of $\lambda_i$ to be
\eqn\valam{ \lambda_i =\prod_{j\neq i} {1-x_i x_j \over x_i-x_j} \qquad i=1,2,...,m }
At these values of the $\lambda$'s, the solution $R_n$ takes
the following particularly simple form: 
\eqn\multisol{ R_n=R {U_n(w_1,...,w_m) U_{n+3}(w_1,...,w_m) \over
U_{n+1}(w_1,...,w_m) U_{n+2}(w_1,...,w_m)} }
where
\eqn\detru{ U_n(w_1,...,w_m)\equiv 
\det\left[ U_{n+2j-2}(w_i) \right]_{1\leq i,j \leq m}}
in terms of Chebyshev polynomials of the first kind expressed at values
$w_i=\sqrt{x_i}+1/\sqrt{x_i}$, $i=1,2,...,m$. 
Eqs.\multisol-\detru\ are proved in Appendix B below, by showing that $u_n^{(m)}\propto
U_n(w_1,...,w_m)$ where the proportionality factors drop out of the ratio giving $R_n$.
The initial conditions $u_{-r}^{(m)}=0$ for $r=1,2,...,2m-1$
translate for the choice \valam\ of the $\lambda$'s into 
the equivalent conditions that $U_{-r}(w_1,...,w_m)=0$.
The latter conditions are clearly satisfied, as direct consequences of the reflection
symmetry of the Chebyshev polynomials, namely that $U_{-n-2}(w)=-U_n(w)$ for
all $n$ (in particular $U_{-1}(w)=0$), and therefore, as a determinant, $U_{-r}(w_1,...,w_m)$ 
always has at least two opposite lines (for even $r$)
or a vanishing one (for odd $r$), provided $r$ stays in the range $1,2,..,2m-1$. 
More precisely, this observation leads to a rank Int$[(1+r)/2]$ for the matrix with
determinant $U_{-r}(w_1,...,w_m)$.
This gives an {\it a posteriori} justification of the choice \valam\ for the
$\lambda$'s.

\newsec{Application: the hard dimer model}

\subsec{The hard dimer model}

\fig{Example (a) of tetravalent two-leg diagram with hard dimers, represented by thickened
edges. The corresponding diagram obtained by shrinking the dimers (b) has both tetravalent
and hexavalent vertices. The correspondence is three-to-one per dimer, as shown.}{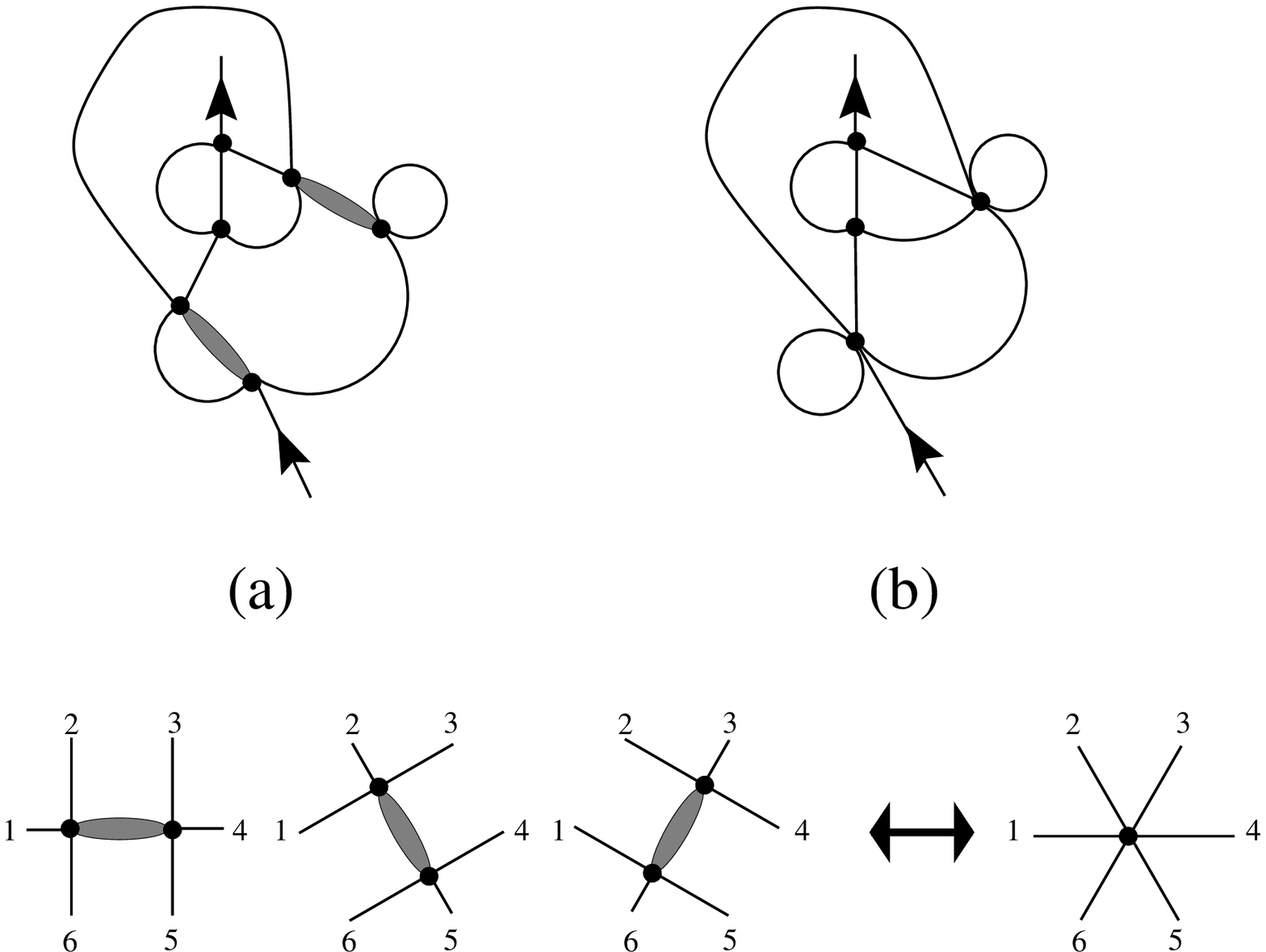}{10.cm}
\figlabel\dimer

As a first application of the exact solution of Sect.5, let us 
consider the so-called hard dimer model on tetravalent planar maps \STAU.
In this model, we enumerate tetravalent maps whose edges may be occupied or not by
a dimer, with the hardness condition that no two adjacent edges may be 
simultaneously occupied by dimers. Assigning as usual a weight
$g$ per vertex and an activity $z$ per dimer, the problem may be reformulated as that
of enumerating maps with both tetra and hexa-valent vertices with a weight
$g_2=g$ per tetravalent vertex and $g_3=3 z g^2$ per hexavalent one. This
is readily seen by shrinking the occupied edges, thus forming hexavalent vertices
out of adjacent pairs of tetravalent ones, as shown in Fig.\dimer. Conversely, any given
hexavalent vertex may be split into a pair of adjacent tetravalent  
ones in three different manners, hence the factor of 3.
This is a particular case of the general model of 
previous section with $m=2$.
Note that
in the hard dimer language,
the notion of geodesic distance must be
understood modulo the restriction that the minimal
path between the two in- and out-coming legs {\it must not} 
cross any edge occupied by a dimer. Indeed, after shrinking
occupied edges it would correspond to going through an hexavalent 
vertex, which is forbidden in the language of graphs with tetra- and hexa-valent
vertices.

In the hard dimer case, eq.\satisR\ reads
\eqn\satidimeR{ R= 1 + 3g R^2+ 30 z g^2 R^3 }
while the recursion \recurela\ reads
\eqn\resuredim{\eqalign{ R_n&= 1 + g R_n(R_{n-1}+R_n+R_{n+1}) 
+3 z g^2 R_n( R_{n-2}R_{n-1}+R_{n-1}^2+R_{n-1}R_{n+1}\cr
&+R_{n+1}^2+R_{n+1}R_{n+2}+R_n(2R_{n-1}+
R_n+2R_{n+1}))\cr}}
The solution now reads
\eqn\soldim{ R_n=R {U_n(w_1,w_2) U_{n+3}(w_1,w_2)\over 
U_{n+1}(w_1,w_2) U_{n+2}(w_1,w_2)}}
where
\eqn\undim{ U_n(w_1,w_2)=U_n(w_1)U_{n+2}(w_2) -U_{n+2}(w_1) U_n(w_2) }
and $w_i=\sqrt{x_i}+1/\sqrt{x_i}$, with $x_1,x_2$ the solutions with modulus
less than one of the characteristic equation
\eqn\cardim{ \chi_2(x)=0 \qquad \chi_2(x)\equiv 1-gR\left(x+{1\over x}+4\right)-3 z g^2R^2
\left(x^2+6 x+16+{6\over x}+{1\over x^2}\right) }
With this we find the initial terms
\eqn\initdim{\eqalign{ R_0&= R{1+(x_1+{1\over x_1})(x_2+{1\over x_2})\over
1+(1+x_1+{1\over x_1})(1+x_2+{1\over x_2})}=R{1-4gR-45zg^2R^2\over
1-3 gR-30 z g^2R^2}\cr
R_1&=R{ 1-7 gR+(11-78z)g^2 R^2+234 z g^3 R^3+1215 z^2 g^4 R^4\over
(1-3 gR-30 z g^2R^2)(1-4gR-45zg^2R^2)}\cr 
R_2&= {1-8 gR +(15-93z)g^2R^2+366z g^3R^3+1836z^2 g^4R^4\over
(1-4gR-45zg^2R^2)(1-7 gR+(11-78z)g^2 R^2+234 z g^3 R^3+1215 z^2 g^4 R^4)}\cr}}
Note that using eq.\satidimeR\ we may recast $R_0=R-g R^3(1+15zgR)$ which 
is in agreement with the general formula \gammatwo\ for the two-leg diagram generating function,
hence corroborating the relation $R_0=\Gamma_2$. 
Finally, we may expand the first few $G_i$'s as
\fig{The tetravalent two-leg diagrams with two inner vertices and one dimer are represented
according to the distance $n$ between the two legs, respectively $n=0$ (a), $n=1$ (b)
and $n=2$ (c). Note that the two first diagrams of (c) have $n=2$ as we forbid crossing the 
dimer.}{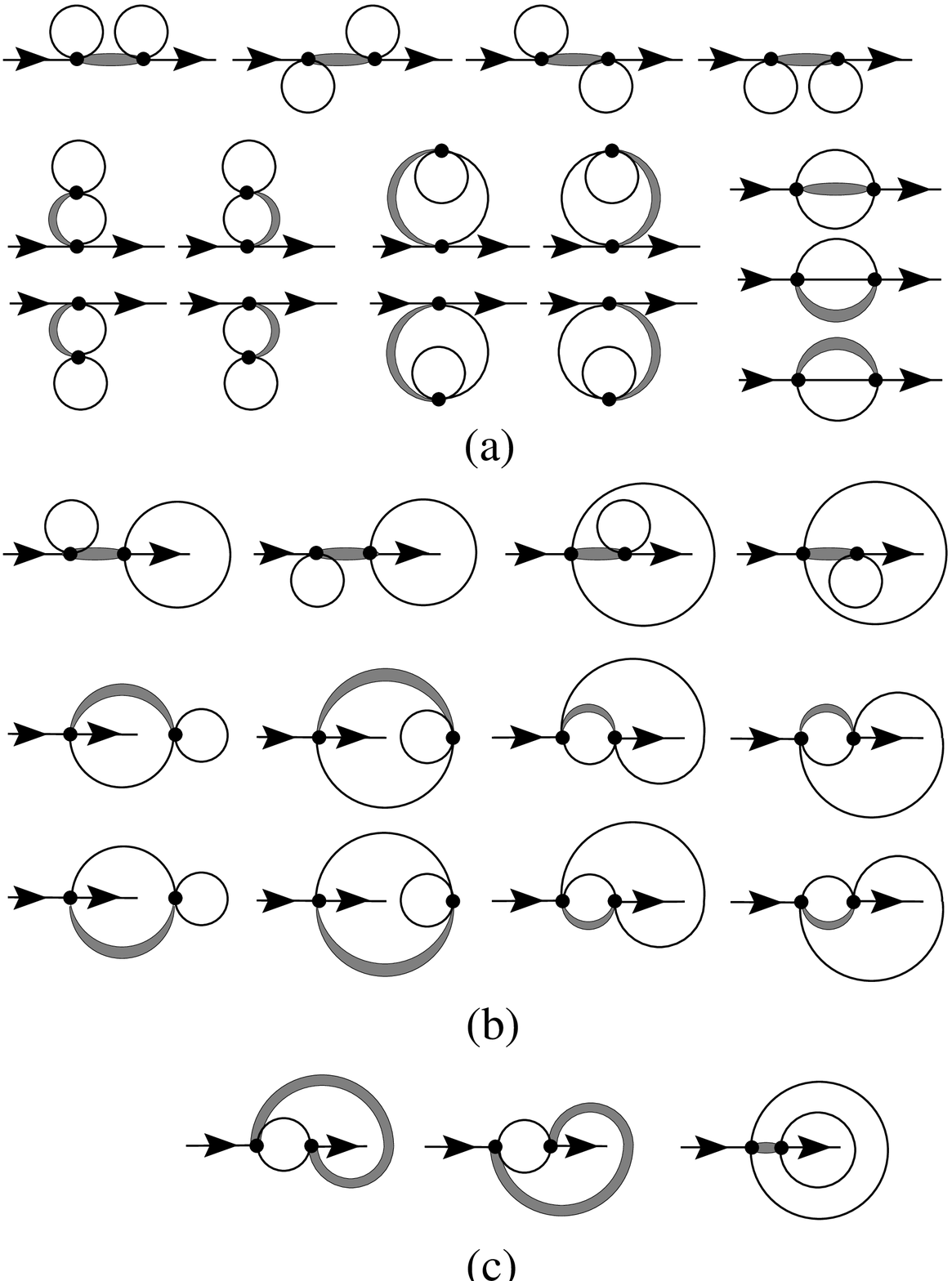}{8.cm}
\figlabel\firdim
\eqn\fewfirg{\eqalign{
G_0(g;z)&=1+2 g+ 3(3+5z)g^2+18(3+10z) g^3+18(21+105 z+50z^2)g^4 +\cdots \cr
G_1(g;z)&=g +4(2+3z)g^2+(65+201z)g^3+2(277+1305z+585z^2)g^4+\cdots \cr
G_2(g;z)&=(1+3z)g^2+3(5+21z)g^3+(179+1005z+531z^2)g^4+\cdots \cr}}
As an illustration, we have represented in Fig.\firdim\
the two-leg diagrams with two
vertices and one hard dimer, corresponding to the terms of order $g^2 z$
in eq.\fewfirg.

\subsec{Tricritical point and fractal dimension}

The hard dimer model possesses a tricritical point characterized as follows.
Introducing the notation
$V=g R$, the eq.\satidimeR\ reads 
\eqn\wg{g=W(V)\equiv V(1-3V-30 z V^2)} 
The critical line of the model is
characterized by the requirement that $W'(V)=0$, leading to the parametric curve 
\eqn\cripar{ g= V\left({2\over 3}-V\right) \qquad z={1-6V \over 90 V^2} }
for $0<V \leq 1/3$. The curve ends at a higher order tricritical point where, in addition
we have $W''(V)=0$, which fixes the critical values
\eqn\critivadim{V_c={1\over 3} \qquad  g_c={1\over 9}\qquad z_c=-{1\over 10}\qquad
R_c={3}} 
We may approach this point by first setting $z=z_c=-1/10$ and letting $g$ approach
$g_c=1/9$, in which case eq.\wg\ simplifies into
\eqn\simpwg{ g=g_c \left(1-\left( {V_c -V \over V_c} \right)^3 \right)} 
We therefore set
\eqn\appro{g=g_c(1-\epsilon^6)\qquad V=V_c(1-\epsilon^2) }
and let $\epsilon \to 0$.
At $z=z_c$, the characteristic equation \cardim\ takes a particularly simple
form when expressed in the variable $y=x+1/x-2$, namely
\eqn\carmieux{ \chi_2(x)=0= {3\over 10} (Vy)^2 - (1-3V) Vy +(1-3V)^2}
We see that at the tricritical point $V=V_c=1/3$, $y=0$ is a double root of
eq.\carmieux, hence
$x=1$ is a quadruple root of the characteristic equation. Moreover, eq.\carmieux\
amounts to the vanishing of a polynomial of degree 2 in the reduced variable $u=y V/(V_c-V)$,
namely $P_2(u)=u^2/30-u/3+1=0$ with the two roots $u_\pm =5\pm i\sqrt{5}$.
When approaching the tricritical point via \appro, we therefore have the solutions
\eqn\ysol{ y_\pm = u_\pm {\epsilon^2 \over 1-\epsilon^2} }
which imply the following leading behavior for the two $x$'s with modulus
less than 1 solving the characteristic equation:
\eqn\twoxsoldim{ x_{1,2}=x_\pm = e^{-\epsilon a_\pm } +O(\epsilon^2)\qquad
a_\pm =\sqrt{5\pm i\sqrt{5}} }
where the square roots are taken with positive real parts.
As before, a sensible scaling limit is obtained by writing $n=r/\epsilon$ in eq.\soldim.
This amounts to introducing a correlation length $\xi =\epsilon^{-1}=((g_c-g)/g_c)^{-1/6}$,
which leads to a critical exponent $\nu=1/6$. 
We may still interpret the quantity $d_F=1/\nu=6$ as a tricritical fractal dimension
as follows. Like in Sect.4.2, we consider the limiting ratio $B_n$ of \behanum\ where 
$R_{n,N}$ now denotes the partition function for two-leg 
diagrams with $N$ vertices, distance at
most $n$ between the two legs and with hard particles with fugacity $z=z_c$.
A straightforward though tedious calculation yields
\eqn\ratiodim{\eqalign{
B_n &= {1\over 16632} {(n+1)(n+6)\over (n+3)(n+4)}\times \cr
&\times  (33264 +74088 n + 63014 n^2+26985 n^3
+6215 n^4+735 n^5+35 n^6)\cr
& \sim {5 \over 2376} n^6\cr}}
in which we read off $d_F=6$. 
 
Further, expanding eq.\soldim\ at order 2 in $\epsilon$,
we obtain the {\it tricritical scaling function} ${\cal F}(r)$ counting surfaces
with two marked points at a geodesic distance larger or equal to $r$:
\eqn\scaldim{\eqalign{
{\cal F}(r)&\equiv \lim_{\epsilon\to 0}{R-R_n\over \epsilon^2 R}=
-2 {d^2 \over dr^2} {\rm Log}\, {\cal W}\left(\sinh \left( a_+{r \over 2} \right), 
\sinh \left( a_-{r \over 2} \right)\right) \cr
&= - {(a_+^2-a_-^2)\left(a_+^2 \sinh^2 \left(a_-{r\over 2}\right)
-a_-^2 \sinh^2 \left(a_+{r\over 2}\right)\right) \over
2\left(a_- \cosh \left(a_-{r\over 2}\right) \sinh \left(a_+ {r\over 2}\right) 
-a_+ \cosh \left(a_+{r\over 2}\right) \sinh \left(a_-{r\over 2}\right)\right)^2} \cr}}
where ${\cal W}(f_1,f_2)$ stands for the Wronskian determinant $f_1 f_2'-f_2f_1'$, and
$a_\pm$ as in eq.\twoxsoldim.
As before, the scaling function ${\cal G}(r)$ counting surfaces with two marked points
at geodesic distance $r$ is obtained by differentiating $-{\cal F}(r)$.

So far we have considered a particular approach to the tricritical point, namely
with a fixed $z=z_c$. Let us now show that the {\it same} exponent $\nu$
is obtained for any other generic approach to the tricritical point and
that moreover 
the {\it same} scaling function is obtained, up to a multiplicative redefinition of $r$.
Indeed, expanding the relation $g=W(V)$ around the critical values as $g=g_c+\delta g$,
$V=V_c+\delta V$, $z=z_c+\delta z$, we obtain at leading order:
\eqn\leaddim{ \delta g+{10\over 9} \delta z= 3 (\delta V)^3 }
Taking a generic approach $\delta g=-g_c \alpha \epsilon^6$, $\delta z=-z_c \beta \epsilon^6$,
we arrive at $\delta V= -V_c (\alpha-\beta)^{1\over 3} \epsilon^2$. 
Accordingly, the characteristic equation \cardim\ reads
\eqn\carred{ \chi_2(x)=0= 3 (\delta V)^2 P\left(-{yV_c\over \delta V }\right)+ O(\epsilon^6) }
which yields the same solution $y_\pm$ as in eq.\ysol\ up to a multiplicative
factor of $(\alpha-\beta)^{1\over 3}$, which in turn leads to the same scaling function
\scaldim\ up to a redefinition $r\to r/(\alpha-\beta)^{1\over 6}$. 

One notable exception is when $\alpha=\beta$, or more precisely when $z_c\delta g=g_c \delta z$ 
in which case the lhs of \leaddim\ vanishes at leading order, 
and we must also consider higher terms
in the expansion, namely of the form $\delta z\delta V$. 
Such a situation occurs in particular if we approach the tricritical point 
{\it along the critical line} \cripar.
In that case, we have $\delta g=-(\delta V)^2$.
Now the appropriate scaling is $\delta g=-g_c \epsilon^4$, and 
$\delta V=-V_c \epsilon^2$. The expansion of the characteristic
equation now leads to another polynomial equation of the form
$Q_2(-yV_c/\delta V))=0$, where $Q_2(u)=u^2/30-u/3$. This gives the solutions 
$x_1=1$ and  
$x_2= e^{-a \epsilon}$, with now $a=\sqrt{10}$.
The appropriate scaling of the distance therefore still reads $n=r/\epsilon$,
and we have an exponent $\nu=1/4$ from the behavior of the 
correlation length $\xi=\epsilon^{-1}=((g_c-g)/g_c)^{-1/4}$. 
This particular approach leads to a different scaling function ${\cal F}_c$ which can be
obtained from the form \scaldim\ by letting $a_+\to 0$ and $a_-=a=\sqrt{10}$, namely
\eqn\scalcridim{{\cal F}_c(r) = -2{d^2\over dr^2} {\rm Log} \left(
{\scriptstyle\sqrt{5\over 2}}\, r\, \cosh\left( {\scriptstyle\sqrt{5\over 2}}\, r\right)-\sinh
\left({\scriptstyle\sqrt{5\over 2}}\, r\right) \right) }

To conclude this section, we note that the generic tricritical
scaling function ${\cal F}(r)$ of eq.\scaldim\ may
be alternatively obtained by deriving the continuum counterpart of eq.\resuredim\ 
yielding the following differential equation for ${\cal F}(r)$
\eqn\painlesix{ {\cal F}^{(4)}(r) -10{\cal F}(r) {\cal F}''(r)-10 {\cal F}''(r)
-5 ({\cal F}'(r))^2+10
({\cal F}(r))^3 +30({\cal F}(r))^2 +30{\cal F}(r) =0}
The scaling function ${\cal F}(r)$ of eq.\scaldim\ is the unique solution
to eq.\painlesix\ such that $\lim_{r\to 0}r^2{\cal F}(r)=6$ 
while $\lim_{r\to \infty} {\cal F}(r)=0$. 
The condition at $r\to 0$ may be seen by writing 
${\cal F}(r)=-2 d^2/dr^2{\rm Log}\, {\cal U}(r)$,
where ${\cal U}(r)$ is the continuum limit of $u_n$ and  
imposing the continuum counterpart of the discrete conditions 
$u_{-1}=u_{-2}=u_{-3}=0$, namely that ${\cal U}(0)={\cal U}'(0)={\cal U}''(0)=0$, 
hence ${\cal U}(r)\propto r^3$, and ${\cal F}(r)\sim 6/r^2$.
In the same spirit, the scaling function ${\cal F}_c$ of eq.\scalcridim\
corresponding to a non-generic approach of the tricritical point along the critical
line is the unique solution with the same boundary conditions as just mentioned
of the truncated differential equation
\eqn\truncdif{ {\cal F}_c^{(4)}(r) -10{\cal F}_c(r) {\cal F}_c''(r)-10 {\cal F}_c''(r)
-5 ({\cal F}_c'(r))^2+10
({\cal F}_c(r))^3 +30({\cal F}_c(r))^2 =0} 

\subsec{Continuum probability distribution for geodesic distances}

As in Sect.4.4, we may translate the above result into an explicit
expression for the probability distribution for geodesic distances in
tricritical surfaces with hard dimers.
Again, we must compute the ratio $R_{n,N}/R_{\infty,N}$, where $R_{n,N}$ now
denotes the generating function for two-leg diagrams with hard dimers at $z=z_c=-1/10$, 
with $N$ inner vertices and legs distant by at most $n$. Starting from the
contour integral \finrn\ applied to the hard dimer problem, we perform the change of variables
$g=W(v)$, with $W$ as in \wg, with the result
\eqn\recchvarW{ R_{n,N}=\oint {dv W'(v)\over 2i\pi W(v)^{N+1}} R_n(v) }
At large $N$, this integral is dominated by the saddle-point $v=V_c=1/3$. 
We now perform the change of variables $v=V_c(1-e^{-i\pi{\rm sgn}(u)/3} |u|/N^{1/3})$
where $u$ is to be integrated on the real line when $N\to\infty$. This prescription
is obtained by deforming the initial contour of integration around the origin of 
the $v$-plane so as to encompass only the image of the pole at $v=0$.
A suitable limit is obtained by further setting $n=\alpha N^{1/6}$, and
using
\eqn\hardrnd{
\eqalign{
R_n(v)&=R\left(1-e^{-i{\pi\over 3}}
{u\over N^{1\over 3}} {\cal F}(\alpha e^{-i{\pi\over 6}}\sqrt{u})+O({1\over N^{2\over 3}})\right)\cr
&={V_c\over g_c}\left(1-e^{-i{\pi\over 3}}{u\over N^{1\over 3}}\left(1+{\cal F}
(\alpha e^{-i{\pi\over 6}}\sqrt{u})\right)+O({1\over N^{2\over 3}})\right)\cr}}
for $u>0$ and ${\cal F}$ as in eq.\scaldim\  and the corresponding counterpart for $u<0$. We finally get 
\eqn\compfinW{
R_{n,N}\sim {9^{N+1}\over \pi N^{4\over 3}}\int_{0}^\infty du
u^3 e^{-u^3}{\rm Re}\left(e^{-i{\pi\over 6}}
(1+{\cal F}(\alpha e^{i{\pi\over 6}}\sqrt{u})\big)\right) }
This leads to the probability for tricritical two-leg diagrams
to have a rescaled geodesic distance less than $\alpha$:
\eqn\probadim{
\Phi(\alpha)=6 {\sqrt{3}\over \Gamma({1\over 3})} \int_0^\infty du u^3e^{-u^3}
{\rm Re}\left(e^{-i{\pi\over 6}}
(1+{\cal F}(\alpha e^{i{\pi\over 6}}\sqrt{u})\big)\right) }
and by differentiating with respect to $\alpha$, to the probability distribution 
$\rho(\alpha)$ for distances equal to $\alpha$.
Note finally that for small $\alpha$, we have $\Phi(\alpha)\sim 5\alpha^6/1188$, in agreement
with eq.\ratiodim, while the large $\alpha$ behavior obeys Fisher's law 
with a decay of the form $\exp(-C \alpha^{6/5})$.

\newsec{General multicritical points: fractal dimensions and scaling functions}

\subsec{Fractal dimensions}

In this section we consider the case of arbitrary $m$, in which case we may reach
an $m$-th order multicritical point. For simplicity, we set $g_1=0$
and we introduce the following notations
\eqn\notaconv{\eqalign{ g_2=g &\qquad g_k=g^{k-1} z_k \ \ ({\rm with}\ z_2=1) \qquad
V=g R \cr
W(V)&= V-\sum_{k=2}^{m+1} z_k {2k-1 \choose k-1} V^k \cr}}
in terms of which eq.\satisR\ reads simply $W(V)=g$. The $m$-th order multicritical
point is obtained by setting $W'=W''=...=W^{(m)}=0$ which implies 
\eqn\valcrig{ z_k^{(c)}=(-1)^k {{1\over m+1}{m+1 \choose k}\over {2k-1\choose k-1}}
\left({6\over m}\right)^{k-1} }
With these, eq.\satisR\ simply reads
\eqn\simpW{ g=W(V)= {V_c^{m+1} -(V_c-V)^{m+1}\over (m+1) V_c^{m}} \qquad V_c={m\over 6}}
while $g_c=W(V_c)=m/(6(m+1))$.
For simplicity, we decide to approach the $m$-th order multicritical point
by first setting all the $z_k$'s to their critical values $z_k^{(c)}$ of
eq.\valcrig, and then letting $g$ approach $g_c$. As already shown in the previous
section, this approach is sufficient to capture the generic multicritical behavior.
As readily seen from eq.\simpW, we must set
\eqn\setcri{ g=g_c(1-\epsilon^{2m+2}) \qquad V=V_c(1-\epsilon^2) }

The characteristic equation \chargen\ takes a particularly simple form when
expressed in the variable $y=x+1/x-2$, namely
\eqn\resimpcar{ \chi_m(x)=0=\sum_{l=0}^m {l!\over (2l+1)!} (Vy)^l {d^{l+1}
W(V) \over dV^{l+1}} }
This is a consequence of the following identity for Chebyshev polynomials
\eqn\idecheby{ \sum_{l=0}^k {2k+1\choose l} U_{2k-2l}(w) =
{(2k+1)!\over k!} \sum_{l=0}^k {l! \over (2l+1)!} {y^l\over (k-l)!} }
where as before $w=\sqrt{x}+1/\sqrt{x}$. This identity is proved by showing 
that both sides satisfy the same recursion relation $\beta_{k+1}-a^2 \beta_k=
{2k+2\choose k+1}$, and have the same initial term $\beta_0=1$.
Eq.\resimpcar\ follows by substituting the identity \idecheby\ into eq.\chargen.  
With the value \simpW\ of $W(V)$, the characteristic equation becomes
\eqn\finchar{\eqalign{ 
\chi_m(x)=0&= \left({V_c-V\over V_c}\right)^m P_m\left({yV\over
V_c-V}\right) \cr 
P_m(u)&=\sum_{l=0}^m  (-u)^l {l! \over (2l+1)!} {m! \over (m-l)!}\cr}}
Approaching the multicritical point as in eq.\setcri,
we get the solutions 
\eqn\solygen{y_i= u_i {\epsilon^2 \over 1-\epsilon^2} }
where the $u_i$, $i=1,2,...,m$ are the distinct roots of $P_m$. In turn, this yields
the $x_i$'s with modulus less than one
\eqn\xigen{ x_i=e^{-\epsilon a_i} +O(\epsilon^2) \qquad a_i=\sqrt{u_i}} 
where the square roots are taken with positive real parts.
Again, to get a sensible limit, we must write $n=r/\epsilon$ in eq.\multisol,
which displays again a correlation length $\xi=\epsilon^{-1}=((g_c-g)/g_c)^{-\nu}$,
with $\nu=1/(2m+2)$.  
This leads to a generalized multicritical fractal dimension
$d_F=2(m+2)$.

\subsec{Scaling functions}

The expansion of eq.\multisol\ at order 2 in $\epsilon$ gives the general
multicritical scaling
function 
\eqn\genscafunc{ {\cal F}(r)=-2 {d^2\over dr^2} {\rm Log}\, 
{\cal W}\left(\sinh\left(a_1{r\over 2}\right), \sinh\left(a_2{r\over 2}\right),...,
\sinh\left(a_m{r\over 2}\right)\right) }
where ${\cal W}(f_1,f_2,...,f_m)$ is the Wronskian determinant $\det\big[ f_i^{(j-1)}
\big]_{1\leq i,j \leq m}$. 
This scaling function may alternatively be obtained as the unique
solution such that $\lim_{r\to 0}r^2{\cal F}(r)=2(2m-1)$
and $\lim_{r\to \infty}{\cal F}(r)=0$ of the differential equation
\eqn\difpain{ {\cal R}_{m+1}[1+{\cal F}]={\cal R}_{m+1}[1] }
where ${\cal R}_p[{\bf u}]$ are the residues of the KdV hierarchy \GD, defined recursively
by ${\cal R}'_{p+1}={\cal R}'''_p/4-{\bf u} {\cal R}'_p-{\bf u}'{\cal R}_p/2$, ${\cal R}_0=1/2$. 

To show eq.\difpain, let us recall that our recursion relation 
\recurela\ coincides with that of the one-matrix model with even potential,
up to a change $1\leftrightarrow n/{\cal N}$. More precisely, writing 
formally eq.\recurela\ as $\Psi_n(\{ R_j\};\{g_k\})=1$, the corresponding
equation in the matrix model reads $\Psi_n(\{ R_j\};\{g_k\})=n/{\cal N}$.
Approaching the $m$-th order critical point with the $g_k$'s given by 
eqs.\notaconv\ and \valcrig, and making the scaling ansatz $g=g_c(1-\epsilon^{2m+2})$,
$R_n=R_c(1-\epsilon^2 {\bf u}(n\epsilon))$, the functional $\Psi_n$ may be expanded at small $\epsilon$
as
\eqn\exppsi{\Psi_n(\{ R_j\};\{g_k\})=1+\epsilon^{2m+2}(1-{\cal S}_{m+1}[{\bf u}])+O(\epsilon^{2m+4})}
where ${\cal S}_{m+1}[{\bf u}]$ is a functional of ${\bf u}$ to be identified. In
the context of the matrix model's double-scaling limit, we write
$\Psi_n(\{ R_j\};\{g_k\})=n/{\cal N}= 1+\epsilon^{2m+2}(1-t)$, where $t$ is the
renormalized cosmological constant, which leads to the equation ${\cal S}_{m+1}[{\bf u}]=t$,
where $\bf u$ is interpreted as the renormalized string susceptibility. From this, 
we identify ${\cal S}_{m+1}\propto {\cal R}_{m+1}$, as is well known from random matrix 
theory \DGZ. Going back to
our problem, the scaling ansatz \scaleq\ with $R=R_c(1-\epsilon^2)$ corresponds
to having ${\bf u}=1+{\cal F}$ and the equation $\Psi_n(\{ R_j\};\{g_k\})=1$
now yields ${\cal S}_{m+1}[1+{\cal F}]=1$. Eq.\difpain\ follows by proportionality 
as we know from eq.\satisR\ that ${\cal F}=0$ must be solution.

As an independent check, we may derive a characteristic equation 
$\pi_{m}(a^2)=0$ by linearizing the continuum equation \difpain\ and looking 
for solutions of the form ${\cal F}(r)=e^{-a r}$. 
Using the explicit value ${\cal R}_p[1]=(-1)^{p} {2p\choose p}/2^{2p+1}$, and the recursion
relation for the KdV residues, we find the
recursion relation $\pi_{p}(a^2)=(a^2-4)\pi_{p-1}(a^2)/4-(-1)^p {2p\choose p}/2^{2p+2}$,
with the initial value $\pi_0=-1/4$. Solving this recursion explicitly, we simply find
that $\pi_m(a^2)= (2m+1)(-1)^{m-1} {2m\choose m}/2^{2m+2} P_m(a^2)$, with $P_m$ as in eq.\finchar, 
hence the characteristic equation coincides with $\chi_m(x)=0$ for 
$x+1/x-2=\epsilon^2 a^2$.

Two final remarks are in order. First, the scaling function \genscafunc\ is the
generic scaling function for the $m$-th order multicritical point. It may
also be obtained by starting with a problem involving vertices up
to valences $2m'+2$, $m'>m$, and approaching any point of the 
$m$-th order critical manifold of dimension $m'-m$, for which the characteristic
equation has $x=1$ as a $2m$-times degenerate zero. 
Conversely, we may approach the $m$-th order multicritical point by first going
to an $m'$-th order multicritical manifold with $m'<m$ (which amounts to having
$x=1$ as a $2m'$-times degenerate zero of the characteristic equation), and then
approaching the $m$-th order multicritical point along this manifold.
The corresponding scaling function is different from the generic
one for the $m$-th order multicritical point and may be obtained as a limit
of the form \genscafunc\ in which 
say $a_1,...,a_{m'}\to 0$ while $a_{m'+1},...,a_m$ tend to the specific
values coming from the restriction of the characteristic equation to the
$m'$-critical manifold. 

\subsec{Probability distribution for geodesic distances}

As in Sects.4.4 and 6.3, we may derive
expressions for the probability distribution of geodesic distances in
multicritical surfaces. Again, we compute the ratio $R_{n,N}/R_{\infty,N}$, 
where $R_{n,N}$ denotes the generating function for two-leg diagrams 
at $z_k=z_k^{(c)}$ as given by eq.\valcrig, 
with $N$ inner vertices and legs at distance at most $n$. In the
the contour integral \finrn\ for the present multicritical problem, 
we perform the change of variables $g=W(v)$, with $W$ as in eq.\notaconv, 
to get, as before
\eqn\recnewchvarW{ R_{n,N}=\oint {dv W'(v)\over 2i\pi W(v)^{N+1}} R_n(v) }
At large $N$, the integral is dominated by the saddle-point $v=V_c=m/6$ and
we perform the change of variables $v=V_c(1-e^{-i{\pi{\rm sgn}(u)\over m+1}} |u|/N^{1\over m+1})$
where $u$ is to be integrated on the real line when $N\to\infty$.
We obtain a suitable limit by further setting $n=\alpha N^{1\over 2(m+1)}$, and
using
\eqn\hardnewrnd{
\eqalign{
R_n(v)&=R\left(1-e^{-i{\pi\over m+1}}
{u\over N^{1\over m+1}} {\cal F}(\alpha e^{-i{\pi\over 2(m+1)}}\sqrt{u})+O({1\over N^{2\over m+1}})\right)\cr
&={V_c\over g_c}\left(1-e^{-i{\pi\over m+1}}{u\over N^{1\over m+1}}\left(1+{\cal F}
(\alpha e^{-i{\pi\over 2(m+1)}}\sqrt{u})\right)+O({1\over N^{2\over m+1}})\right)\cr}}
for $u>0$ with ${\cal F}$ as in eq.\genscafunc\ and the corresponding counterpart for $u<0$. 
We finally get
\eqn\compfinewW{
R_{n,N}\sim {\left(6{m+1\over m}\right)^{N}(m+1)^2\over \pi N^{m+2\over m+1}}\int_{0}^\infty du
u^{m+1} e^{-u^{m+1}}{\rm Re}\left(e^{-i{\pi(m-1)\over 2(m+1)}}
(1+{\cal F}(\alpha e^{i{\pi\over 2(m+1)}}\sqrt{u})\big)\right) }
This gives the probability distribution for rescaled geodesic distances at most $\alpha$
\eqn\probanewdim{
\Phi(\alpha)={(m+1)^2\over\cos\left({\pi(m-1)\over 2(m+1)}\right) \Gamma({1\over m+1})} 
\int_{0}^\infty du
u^{m+1} e^{-u^{m+1}}{\rm Re}\left(e^{-i{\pi(m-1)\over 2(m+1)}}
(1+{\cal F}(\alpha e^{i{\pi\over 2(m+1)}}\sqrt{u})\big)\right) }
and that for rescaled geodesic distances equal to 
$\alpha$ by differentiation. As expected, we have $\Phi(\alpha)\propto \alpha^{2(m+1)}$ at
small $\alpha$ while $\rho(\alpha)$ decays for large $\alpha$
as $\exp(-C \alpha^{2(m+1)/(2m+1)})$.

\newsec{Discussion and conclusion}

In this paper, we have found explicit expressions for partition
functions of planar maps with arbitrary even valences and
with two marked points at a specified geodesic distance.  
{}From these we have extracted various scaling functions and critical
exponents corresponding to various approaches to the different 
critical and multicritical points. We also obtained various probability
distributions for distances in maps of fixed but large number of 
vertices. These results rely on a purely combinatorial technique based 
on bijections with trees. 

\subsec{Relation to matrix models and integrability}

The results obtained here cannot apparently be reached by the more 
standard matrix integral approach, which naturally deals with the genus
of the graphs rather than with geodesic distances. However, 
we observe strong analogies between the two problems, as apparent 
from the structure of our main recursion relation. Even more
strikingly, in the continuum limit of the $m$-critical model, we
found a differential equation of the form 
${\cal R}_{m+1}[{\bf u}]={\cal R}_{m+1}[1]$ with ${\bf u}=1+{\cal F}$, to be compared
with the generalized KdV equation ${\cal R}_{m+1}[{\bf u}]= {\rm const.}\, t$ 
for the all-genus string
susceptibility ${\bf u}(t)$ where $t$ plays the role of a renormalized 
cosmological constant in the double-scaling limit \DGZ. 
This also amounts to
having replaced the usual string equation $[P,Q]=1$, $Q=d^2-{\bf u}$,
$d=d/dt$, and $P$
a differential operator of degree $2m+1$, by simply $[P,Q]=0$.
This modification is what makes the problem exactly solvable both
in the discrete and continuum versions. In particular, we note
the appearance of soliton-like expressions for our solutions, 
which are the sign of an underlying integrable structure.

We expect that this scheme also applies to the general case of a
$c(p,q)$ CFT coupled to 2D quantum gravity, namely that the
fundamental two-point scaling functions of the geodesic distance
are still determined 
by differential equations of the form $[P,Q]=0$, where
$P$ and $Q$ are differential operators of respective degrees $p$ and $q$.

\subsec{Other two-point functions}

The continuum $m$-multicritical two-point function that we have derived
was obtained as the scaling limit of the exact discrete solution. 
We expect on general
grounds that it corresponds to the two-point correlation of the {\it most relevant}
operator $\phi_1$ in the corresponding $c(2,2m+1)$ conformal field theory 
coupled to 2D quantum gravity.
It is clear that many more quantities may be derived from the 
general solution \mxsol. 
In the language of quantum gravity, this includes in particular correlations
of the form $\langle \phi_1(0)\phi_j(r)\rangle$
involving less relevant operators $\phi_j$, $j=2,3,...,m$
as functions of the geodesic distance $r$. 
In the language of this paper, natural discrete candidates for such
correlators are the generating functions $G_n^{(k)}$
for two-leg diagrams with legs distant by $n$ 
and such that the outcoming leg is attached to a $2k$-valent vertex.
In this paper we have considered $G_n=\delta_{n,0}+\sum G_n^{(k)}$,
a particular linear combination of the $G_n^{(k)}$. 
In the continuum limit, this combination has a non-vanishing overlap with the most relevant
operator $\phi_1$ of the theory, hence we may interpret our result as computing
$\langle \phi_1(0)\phi_1(r) \rangle \propto {\cal G}(r)$. We expect that by considering
other fine-tuned linear
combinations of the $G_n^{(k)}$, we will obtain higher order terms in $\epsilon$,
a term of order $\epsilon^{2j+1}$  
corresponding to a two-point function $\langle \phi_1(0)\phi_j(r)\rangle$. 
Indeed, the above linear combinations amount to having inserted some operator
$\phi_j$ at the vertex adjacent to the outcoming leg, while the other leg
remains untouched, and therefore still carries $\phi_1$.
To compute these two-point functions, we observe that 
$R_n^{(k)}\equiv \sum_{j=0}^n G_j^{(k)}=g_k(Q^{2k-1})_{n-1,n}$,
with notations as in eqs.\opq-\recumieux. These are the same polynomials of the $R_j$'s
as those appearing in the recursion relation for the one-matrix model, and we may
therefore apply the known results
on the double-scaling limit in relation with KdV flows to identify 
\eqn\twoptother{ \langle \phi_1(0)\phi_j(r)\rangle \propto -{d \over dr} {\cal R}_j[1+{\cal F}]}
for $j=1,2,...,m$, 
with $\cal F$ solving eq.\difpain.

These correlators also allow for an
explicit exploration of the general problem of fractal dimensions,
a subtle question already addressed in Refs.[\xref\AJW, \xref\KY].

\subsec{Generalizations} 

\fig{Example (a) of a two-leg $3$-constellation with geodesic distance $1$, and (b)
the corresponding blossom tree (see Ref.\BMS\ for precise definitions and
characterizations).}{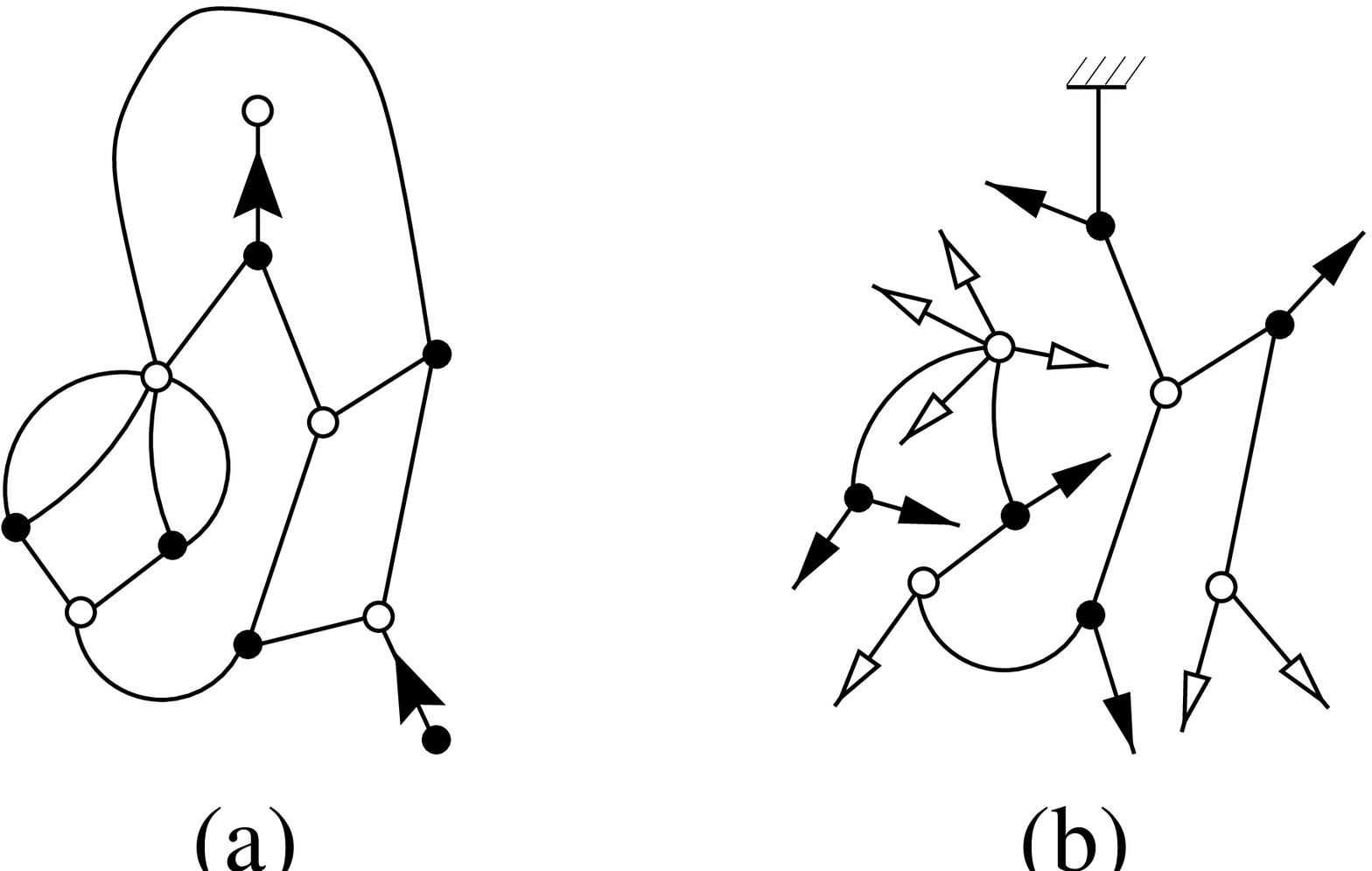}{10.cm}
\figlabel\constella

We may adapt the techniques of this paper so as
to address similar questions in more involved graph enumeration
problems. Indeed, the case of planar maps with vertices
of even valence that we considered here is a particular case of a 
more general class
of planar maps called $p$-constellations. These are bipartite planar
maps with say black and white vertices, such that all black vertices are
$p$-valent, while the white ones have valences $kp$, $k=1,2,3...$.
For $p=2$, the black (two-valent) vertices may be wiped out while
the white ones have arbitrary even valences, which is the case studied
in this paper.
For general $p\geq 3$, we may still consider two-leg diagrams with a univalent
black vertex (in-coming leg) and a univalent white one (out-coming leg),
not necessarily adjacent to the same face (see Fig.\constella\ for
an illustration). The
geodesic distance between the legs is the minimal length of paths
connecting them, with the constraint that one may only cross edges with the white
vertex on the right and therefore the black vertex on the left, when going from the face
adjacent to the in-coming leg to that adjacent to the out-coming one. 
These two-leg diagrams are in one-to-one correspondence with suitable bipartite
blossom trees as shown in Ref.\BMS. Using this equivalence, and limiting ourselves to
white vertex valences up to $(m+1)p$ for some fixed $m$ we have shown 
that the generating function $R_n$ for two-leg
diagrams with legs distant by at most $n$ satisfies a 
polynomial recursion relation, which allows to determine them exactly.
Remarkably, we find that the generic form for $R_n$ is again very similar to that
of \mxsol, except for a simple shift in the indices of $u_n^{(N)}$, namely
\eqn\constu{ R_n=R {u_n^{(N)} u_{n+p+1}^{(N)}\over u_{n+1}^{(N)} u_{n+p}^{(N)}} }
where $R=\lim_{n\to \infty} R_n$ is a solution of a polynomial equation of
degree $(m+1)(p-1)$, $u_n^{(N)}$ is a $N$-soliton solution to KP of the form
\jimiw\ with $N=(m+1)(p-1)-1$, $e^{\eta_i}=-x_i^{n+N} \lambda_i$, $x_i$ being
the solutions with modulus less than 1 of the characteristic equation 
associated to the recursion relation (and of degree $N$ in $x+1/x$), and finally
$p_i=x_i+x_i^2+...+x_i^{p-1}$, while $q_i=1/x_i+1/x_i^2+...+1/x_i^{p-1}$.  
Again, the integration constants $\lambda_i$ are determined by the initial conditions
$u_{-i}=0$, $i=1,2,...,(2N-1)$, which in turn fully specifies the solution $R_n$. 
Apart from these complications, the strategy of the present paper may be applied
to these more involved cases of $p$-constellations. In particular, we have checked
that we recover the expected generating function $\Gamma_2=R_0$ previously obtained
in Refs.[\xref\BFG-\xref\NOUSHARD]. However, we do not expect any
new critical behavior or scaling functions other than that already obtained in
this paper for $p=2$. 

More promising, the same techniques can be applied to models of graphs with
matter degrees of freedom such as the Ising model, whose study is under 
way \NOUSISING, and where new scaling functions appear.  

\subsec{Connection to ISE}

Let us finally discuss the relation between the present work and the other
combinatorial approach of Ref.\CS\ for the tetravalent case. In this
reference, a different bijection between tetravalent rooted maps and 
labeled trees is used, where the labels precisely encode the geodesic
distance from the root. Representing both the underlying tree and its
labels by two discrete 1D random walks, the problem translates into
a discrete version of the so-called Brownian snake, which makes the
connection between geodesic distances on tetravalent 
planar maps and random variables of the Integrated SuperBrownian Excursion (ISE).
In particular, the continuum distribution $\Phi(\alpha)$ for the simple
critical case $m=1$ coincides
with the distribution for the maximum of the ISE,  
whose suitably defined Laplace transform  was obtained in Ref.\DEL\ and
coincides with ${\cal F}$. Beyond this $m=1$ case, 
it would be interesting to understand in ISE terms the counterpart
of our multicritical results for arbitrary $m$. In particular,
eq.\probanewdim\ may be inverted as
\eqn\invphif{{\cal F}(\alpha)={1\over (m+1)\Gamma\left({m\over m+1}\right)}\int_0^\infty{du \over u^{m+2\over m+1}}e^{-u}
\left(1-\Phi\left({\alpha\over u^{1\over 2(m+1)}}\right)\right)}
which displays the suitably generalized Laplace transform to be used 
in this case. In the ISE
language, this suggests to consider excursions with ``duration" $u$ distributed
according to a suitably generalized Ito law $P(u)\propto 1/u^{m+2\over m+1}$. 
An example of the discrete
version of such ``multicritical excursions" was considered in the context 
of $(1+1)$-D Lorentzian gravity in Ref.\DGL\ in the form of walks 
with steps of heights $1,2,\cdots,m$ with respective weights 
$z_1,\cdots,z_m$ which must be fine-tuned to some multicritical 
values. 
Understanding such a connection would provide a direct link between the so-called
Lorentzian and Euclidean two-dimensional quantum gravities.

\noindent{\bf Acknowledgments}

We thank F. David, J.-F. Delmas and G. Schaeffer for useful discussions partly while
attending the semester ``Geometry and statistics of random growth" held at the
Institut Henri Poincar\'e, Paris, January-March 2003.  We also thank D. Bernard, 
R. Conte and M. Musette for help with references on solitons.

\bigskip

\appendix{A}{Proof of the determinant form \detergent\ for $u_n^{(m)}$ of eq.\mxsol.}

Eq.\detergent\ is proved by expanding the determinant as
\eqn\reduk{\eqalign{ \det \, M_n^{(m)}&= \sum_{\sigma \in S_m} {\rm sgn}(\sigma) \prod_{i=1}^m
\left(\delta_{i,\sigma(i)} -
\lambda_i x_i^{n+m}{1-x_i^2\over 1-x_ix_{\sigma(i)}}\right)\cr
&= \sum_{\sigma \in S_m} {\rm sgn}(\sigma)
\sum_{l=0}^m \sum_{1\leq i_1<...<i_{l}\leq m} \prod_{j\notin \{i_1,...,i_l\}}
\delta_{j,\sigma(j)} \prod_{i\in \{i_1,...,i_l\}} \big(-\lambda_{i} x_i^{n+m}
{1-x_i^2\over 1-x_ix_{\sigma(i)}}\big) \cr
&=\sum_{l=0}^m (-1)^l \sum_{1\leq i_1<...<i_{l}\leq m} \left(\prod_{r=1}^l
\lambda_{i_r} x_{i_r}^{n+m}\right) \  \det\left[ {1-x_{i_s}^2\over 1-x_{i_s} x_{i_t}}
\right]_{1\leq s,t \leq l} \cr}}
The identification with the form \mxsol\ for $u_n^{(m)}$ is completed by computing
the determinant in the last line of \reduk\ as
\eqn\compek{
\det\left[ {1-y_s^2\over 1-y_s y_t} \right]_{1\leq s,t \leq l}
=\prod_{1\leq s<t\leq l} \left({y_s-y_t\over 1-y_s y_t}\right)^2}
where $y_s\equiv x_{i_s}$, obtained
by direct application of the Cauchy determinant formula
\eqn\cauch{ \det\left[{1\over y_s+z_t}\right]_{1\leq s,t \leq l}=
{\displaystyle \prod_{1\leq s<t\leq l} (y_s-y_t)(z_s-z_t) \over
\displaystyle \prod_{1\leq s,t\leq l} (y_s+z_t)} }
upon taking $z_s=-1/y_s$.

\appendix{B}{Proof of the determinant form \multisol-\detru.}

Let us start from the determinant formula \detergent\ in which we substitute
the values \valam\ of the $\lambda_i$: 
\eqn\gentdeter{\eqalign{ &u_n^{(m)}=\det\  M_n^{(m)} \cr
&\big[M_n^{(m)}\big]_{i,j}=\delta_{i,j} - x_i^{n+m}
{\displaystyle\prod_{l\neq j} {1-x_i x_l} \over 
\displaystyle\prod_{l\neq i} x_i-x_l} \qquad i,j=1,2,...,m \cr} }
We now multiply $M_n^{(m)}$ to the right by the matrix $P^{(m)}$ with entries
\eqn\entryp{ \left[ P^{(m)}\right]_{i,j}={x_i^{m-j}\over 
\displaystyle\prod_{l\neq i} x_i-x_l} }
This gives the general term
\eqn\geter{ \left[ M_n^{(m)} P^{(m)} \right]_{i,j}= {x_i^{m-j}\over 
\displaystyle\prod_{l\neq i} x_i-x_l} \left( 1- x_i^{n+j} \sum_{k=1}^m
x_k^{1-j} \prod_{l\neq k} {1-x_i x_l\over 1-{x_l \over x_k}} \right) }
Defining 
\eqn\phidef{ \varphi_j(x)= \sum_{k=1}^m
x_k^{1-j} \prod_{l\neq k} {1-x x_l\over 1-{x_l \over x_k}}}
we see that $\varphi(x)$ is the Lagrange interpolation polynomial
of degree less or equal to $m-1$, that takes the values $\varphi_j(1/x_k)=x_k^{1-j}$
at the $m$ distinct points $x=1/x_k$,
for $k=1,2,...,m$. We conclude that
$\varphi_j(x)=x^{j-1}$ for $j=1,2,...,m$.
This allows to finally rewrite \geter\ as
\eqn\lutint{\eqalign{ \left[ M_n^{(m)} P^{(m)} \right]_{i,j}&={x_i^{m+{n-1\over 2}}\over 
\displaystyle\prod_{l\neq i} x_i-x_l} \left({1\over x_i^{{n-1\over 2}+j}}- 
x_i^{{n-1\over 2}+j} \right) \cr
&= {x_i^{m+{n-1\over 2}}\over 
\displaystyle\prod_{l\neq i} x_i-x_l}  \left({1\over \sqrt{x_i}}-\sqrt{x_i}\right)
U_{n+2j-2}(w_i) \cr}}
with $w_i=\sqrt{x_i}+1/\sqrt{x_i}$ as before.
Taking the determinant, we arrive at
\eqn\arrivons{ u_n^{(m)}= {1\over \det P^{(m)}} \prod_{i=1}^m \left({x_i^{m+{n-1\over 2}}
({1\over \sqrt{x_i}}-\sqrt{x_i})\over 
\displaystyle\prod_{l\neq i} x_i-x_l}\right) 
\times \det \left[U_{n+2j-2}(w_i)\right]_{1\leq i,j\leq m}}
which immediately leads to eqs.\multisol-\detru, by noting that all the prefactors drop out
of the ratio.

\listrefs
\end
That the values \valam\ actually ensure these initial conditions may be checked
directly by exhibiting null vectors for the corresponding matrix
$M_n^{(m)}$, $n=-1,-2,...,-(2m-1)$.
Let indeed
\eqn\nulvec{ \big[v_n^{(m)}\big]_i= {f_n^{(m)}(x_i)
\over \displaystyle\prod_{l\neq i} (x_i-x_l)} }
we see that $v_n^{(m)}$ is a null vector for $M_n^{(m)}$ if
\eqn\sinvec{ {f_n^{(m)}(x_i) \over x_i^{n+m}} =\sum_{j=1}^m {f_n^{(m)}(x_j) \over x_j^{m-1}}
\displaystyle \prod_{l\neq j} {1-x_i x_l \over 1-{x_l\over x_j}} }
Let us look for functions $f_n^{(m)}$ such that
\eqn\moinsfort{ {f_n^{(m)}(x) \over x^{n+m}} =\sum_{j=1}^m {f_n^{(m)}(x_j) \over x_j^{m-1}}
\prod_{l\neq j} {1-x x_l \over 1-{x_l\over x_j}} }
The rhs is nothing but the Lagrange interpolation polynomial $P$ of degree $m-1$, such
that $P(1/x_i)={f_n^{(m)}(x_i) / x_i^{m-1}}$, itself equal to the lhs ${f_n^{(m)}(x)/
x^{n+m}}$. We deduce that $f_n^{(m)}$ is a polynomial with degree at most $m-1$ and valuation at
least
$n+m$, and satisfying the reciprocality property
\eqn\recipro{ f_n^{(m)}(x)= x^{2m+n-1} f_n^{(m)}\left({1\over x}\right) }
In other words, the coefficients of $x^s$ and $x^{2m+n-1-s}$ must be equal in $f$.
{}From the condition on the valuation and degree, such polynomials exist iff
$-1 \geq n \geq -(2m-1)$, which leads to Int$[(1-n)/2]$ independent solutions.
In these cases, we may pick for instance
$f_{-(2r-1)}^{(m)}(x)= x^{m-r}$, $r=1,2,...,m$  and $f_{-2r}^{(m)}(x)=x^{m-r-1}(1+x)$,
$r=1,2,...,m-1$.
Any such polynomial leads to a non-vanishing null vector via eq.\nulvec.
This shows that $u_{-1}^{(m)}=...=u_{-(2m-1)}^{(m)}=0$.